\documentclass{aa}

\usepackage{graphicx}
\usepackage[varg]{txfonts}
\usepackage{natbib}
\bibpunct{(}{)}{;}{a}{}{,}
\usepackage{array}
\usepackage{xspace}
\usepackage{lscape}
\usepackage{url}
\usepackage{arydshln}
\usepackage[hyperfootnotes=false, linktocpage=true, breaklinks=true, colorlinks=true, linkcolor=blue, citecolor=blue, urlcolor=blue]{hyperref}
\usepackage[all]{hypcap}
\usepackage{longtable}
\usepackage{afterpage}

\newcommand{\FeKa}{Fe K\ensuremath{\alpha}\xspace}

\newcommand{\kms}{\ensuremath{\mathrm{km\ s^{-1}}}\xspace}
\newcommand{\NH}{\ensuremath{N_{\mathrm{H}}}\xspace}

\newcommand{\xmm}{{\it XMM-Newton}\xspace}
\newcommand{\chandra}{{\it Chandra}\xspace}
\newcommand{\swift}{{\it Swift}\xspace}

\newcommand{\spitzer}{{\it Spitzer}\xspace}
\newcommand{\hst}{{HST}\xspace}
\newcommand{\fuse}{{\it FUSE}\xspace}

\newcommand{\mrk}{{Mrk~509}\xspace}
\newcommand{\ngc}{{NGC~7469}\xspace}

\newcommand{\nustar}{{\it NuSTAR}\xspace}

\newcommand{\ergcm}{{\ensuremath{\rm{erg\ cm}^{-2}\ \rm{s}^{-1}\ {\AA}^{-1}}}\xspace}
\newcommand{\ergflux}{{\ensuremath{\rm{erg\ cm}^{-2}\ \rm{s}^{-1}}}\xspace}
\newcommand{\ergs}{{\ensuremath{\rm{erg\ s}^{-1}}}\xspace}
\newcommand{\cm}{{\ensuremath{\rm{cm}^{-2}}}\xspace}

\newcommand{\xstar}{{\tt XSTAR}\xspace}
\newcommand{\cloudy}{{\tt Cloudy}\xspace}
\newcommand{\spex}{\xspace{\tt SPEX}\xspace}
\newcommand{\pion}{\xspace{\tt pion}\xspace}
\newcommand{\cie}{\xspace{\tt cie}\xspace}

\newcommand{\chart}{\xspace{\tt ChaRT}\xspace}
\newcommand{\marx}{\xspace{\tt MARX}\xspace}

\mathchardef\mhyphen="2D
\begin{document}

\title{Multi-wavelength campaign on NGC 7469}
\subtitle{III. Spectral energy distribution and the AGN wind photoionisation modelling, plus detection of diffuse X-rays from the starburst with \textit{\textbf{Chandra}} HETGS}

\author{
M. Mehdipour \inst{1}
\and
J.S. Kaastra \inst{1,2}
\and 
E. Costantini \inst{1}
\and
E. Behar \inst{3}
\and
G.A. Kriss \inst{4}
\and
S. Bianchi \inst{5}
\and
G. Branduardi-Raymont \inst{6}
\and
\newline
M. Cappi \inst{7}
\and
J. Ebrero \inst{8}
\and
L. Di Gesu \inst{9}
\and
S. Kaspi \inst{3}
\and
J. Mao \inst{1,2}
\and
B. De Marco \inst{10}
\and
R. Middei \inst{5}
\and
U. Peretz \inst{3}
\and
P.-O. Petrucci \inst{11}
\and
\newline
G. Ponti \inst{12}
\and
F. Ursini \inst{7}
}

\institute{
SRON Netherlands Institute for Space Research, Sorbonnelaan 2, 3584 CA Utrecht, the Netherlands\\ \email{M.Mehdipour@sron.nl}
\and
Leiden Observatory, Leiden University, PO Box 9513, 2300 RA Leiden, the Netherlands
\and
Department of Physics, Technion-Israel Institute of Technology, 32000 Haifa, Israel
\and
Space Telescope Science Institute, 3700 San Martin Drive, Baltimore, MD 21218, USA
\and
Dipartimento di Matematica e Fisica, Universit\`{a} degli Studi Roma Tre, via della Vasca Navale 84, 00146 Roma, Italy
\and
Mullard Space Science Laboratory, University College London, Holmbury St. Mary, Dorking, Surrey, RH5 6NT, UK
\and
INAF-IASF Bologna, Via Gobetti 101, I-40129 Bologna, Italy
\and
European Space Astronomy Centre, P.O. Box 78, E-28691 Villanueva de la Ca\~{n}ada, Madrid, Spain
\and
Department of Astronomy, University of Geneva, 16 Ch. d'Ecogia, 1290 Versoix, Switzerland
\and
Nicolaus Copernicus Astronomical Center, Polish Academy of Sciences, Bartycka 18, PL-00-716 Warsaw, Poland
\and
Univ. Grenoble Alpes, CNRS, IPAG, 38000 Grenoble, France
\and
Max Planck Institute fur Extraterrestriche Physik, 85748, Garching, Germany
}

\date{Received 8 January 2018 / Accepted 21 March 2018}
\abstract
{
We investigate the physical structure of the AGN wind in the Seyfert-1 galaxy \ngc through high-resolution X-ray spectroscopy with \chandra HETGS and photoionisation modelling. Contemporaneous data from \chandra, \hst, and \swift are used to model the optical-UV-X-ray continuum and determine the spectral energy distribution (SED) at two epochs, 13 years apart. For our investigation we use new observations taken in December 2015--January 2016, and historical ones taken in December 2002. We study the impact of a change in the SED shape, seen between the two epochs, on the photoionisation of the wind. The HETGS spectroscopy shows that the AGN wind in \ngc consists of four ionisation components, with their outflow velocities ranging from $-400$ to $-1800$~\kms. From our modelling we find that the change in the ionising continuum shape between the two epochs results in some variation in the ionisation state of the wind components. However, for the main ions detected in X-rays, the sum of their column densities over all four components, remains in practice unchanged. For two of the four components, which are found to be thermally unstable in both epochs, we obtain ${2 < r < 31}$~pc and ${12 < r < 29}$~pc using the cooling and recombination timescales. For the other two thermally stable components, we obtain ${r < 31}$~pc and ${r < 80}$~pc from the recombination timescale. The results of our photoionisation modelling and thermal stability analysis suggest that the absorber components in \ngc are consistent with being a thermally-driven wind from the AGN torus. Finally, from analysis of the zeroth-order ACIS/HETG data, we discover that the X-ray emission between 0.2--1 keV is spatially extended over 1.5--12$\arcsec$. This diffuse soft X-ray emission is explained by coronal emission from the nuclear starburst ring in \ngc. The star formation rate inferred from this diffuse soft X-ray emission is consistent with those found by far-infrared studies of \ngc.
}
\keywords{X-rays: galaxies -- galaxies: active -- galaxies: Seyfert -- galaxies: individual: NGC 7469 -- techniques: spectroscopic}
\authorrunning{M. Mehdipour et al.}
\titlerunning{Multi-wavelength campaign on NGC 7469. III.}
\maketitle
\section{Introduction}

Outflows of gas from active galactic nuclei (AGN) couple the supermassive black holes (SMBHs) to their environment. The observed relations between SMBHs and their host galaxies, such as the M-$\sigma$ relation \citep{Ferr00}, indicate that SMBHs and their host galaxies are likely co-evolved through a feedback mechanism. The AGN outflows may play a key role in this co-evolution as they can significantly impact star formation (e.g. \citealt{Silk98,King10}), chemical enrichment of the surrounding intergalactic medium (e.g. \citealt{Oppen06}), and the cooling flows at the core of galaxy clusters (e.g. \citealt{Ciott01}).

Winds of photoionised gas are an essential component of outflows in AGN. These winds, called warm absorbers (WA, hereafter), imprint their absorption lines on the AGN continuum, providing us with important diagnostics. High-resolution UV and X-ray spectroscopy, currently possible by the grating spectrometers of \hst, \chandra, and \xmm, enables us to study these winds in bright AGN. Spectroscopic studies show that AGN winds often consist of multiple ionisation and velocity components (e.g. \citealt{Cren99,Blu05,McKe07}). In some cases, winds have extreme column densities that obscure the central X-ray source (e.g. \citealt{Kaas14,Mehd17}), or are highly ionised with relativistic velocities (e.g. \citealt{Reev09}). There are however significant gaps in our understanding of the AGN winds, in particular with respect to their origin, launching mechanism, and their impact on their environment.

As AGN outflows are ultimately powered and driven by energy released from the accretion process, their properties are expected to be related to the physical conditions of the accretion disk and its associated radiation components. However, it is ambiguous what physical factors govern the launch of outflows, and what is the dependence on the properties of accretion onto SMBHs. It has been suggested that ionised outflows in AGN could be thermally-driven winds from the AGN torus (e.g. \citealt{Krol01}), or that they originate as radiatively driven (e.g. \citealt{Prog00}), or magnetohydrodynamic (e.g. \citealt{Fuku10}) winds from the accretion disk. Determining the physical structure and origin of the outflows, and understanding their role in shaping AGN spectra and variability, are crucial requirements for a general characterisation of the outflows and advancing our knowledge of AGN/galaxy evolution.

Photoionisation modelling is a powerful way to understand the nature and origin of AGN winds. Photoionisation codes, namely \cloudy \citep{Ferl17}, \spex \citep{Kaa96}, and \xstar \citep{Kall01}, enable us to interpret the high-resolution spectra, and thus play a crucial role in our understanding of AGN winds. Photoionisation calculations are strongly dependent on our knowledge of the SED of the central ionising source (see e.g. \citealt{Chak12,Meh16b}). In particular, the extreme ultraviolet (EUV) continuum is a highly significant part of the SED as it strongly influences the ionisation and thermal state of the wind. However, EUV photons are in practice not detectable as they are diminished by Galactic absorption. Therefore, knowledge of the formation of the accretion-powered optical-UV-X-ray continuum components is important for constructing realistic SEDs, and carrying out accurate photoionisation modelling. Robust parameterisation of the wind components through photoionisation modelling and spectroscopy, would enable us to map the ionisation, dynamical, and chemical structure of the wind in AGN. The derived physical properties of the wind, its absorption measure distribution (AMD, \citealt{Holc07}, \citealt{Beha09}), and whether its components are thermally stable and in pressure equilibrium or not, would provide important clues for linking the observations to the appropriate theoretical models for formation and driving of AGN winds. Such a characterisation of AGN outflows is required in order to understand their role and impact in galaxy evolution in the universe. The Seyfert-type AGN in the nearby universe, which are sufficiently bright for high-resolution UV and X-ray spectroscopy with the existing observatories, are currently the most suitable laboratories to carry out such detailed investigations into AGN winds.

\object{NGC 7469} is a Seyfert-1 AGN at the redshift of 0.016268 \citep{Spri05} with a SMBH mass of about ${1 \times 10^{7}}$~$M_{\odot}$ \citep{Pete14}. The host galaxy of \ngc is classified as a luminous infrared galaxy (LIRG). The nucleus of \ngc is surrounded by a well-known starburst ring, which has been detected in multiple energy bands from radio to optical/UV (e.g. \citealt{Wils91,Maud94}). Previous high-resolution UV and X-ray spectroscopic studies reported about the presence of a nuclear wind in this AGN. The UV studies with \fuse and \hst STIS found two distinct absorption components, outflowing with velocities of $-570$ and $-1900$~\kms \citep{Kris03,Scot05}. The historical high-resolution X-ray studies, using a 40~ks \xmm observation \citep{Blu03} and a 150~ks \chandra HETGS observation \citep{Scot05}, found multiple absorption components. However, these archival X-ray data were limited to constrain the X-ray properties of the WA. In 2015 we carried out an extensive multi-wavelength campaign on \ngc \citep{Beh17}, using \xmm, \nustar, \swift, \chandra, and \hst. The analysis of the stacked 2015 \xmm RGS spectrum is reported in \citet{Beh17}, where various components of the warm absorber/emitter were determined. In addition to spectral lines from the AGN wind, \ion{Fe}{xvii} emission lines from a collisionally-ionised component were also found, which is likely associated to the starburst activity of \ngc. Furthermore, variability in the ionic column density of the warm absorber between the individual 2015 RGS spectra are investigated in \citet{Pere18}, where lines from individual ions are independently measured. No significant changes in the ionic column densities were found among the 2015 observations. 

This paper is the primary publication of our new 2015 \chandra HETGS observations. We determine the SED of \ngc in 2002 and 2015 from optical to hard X-rays using contemporaneous \chandra, \hst and \swift observations. These observations and the processing of their data are described in Sect. \ref{data_sect}. We take into account the contribution of all non-intrinsic emission and absorption processes in our line of sight towards the central engine of \ngc in Sect. \ref{sed_sect}, in order to determine the underlying AGN continuum. We then carry out photoionisation modelling and HETGS spectroscopy of the wind in Sect. \ref{model_sect}, where we investigate the long-term variability of the AGN wind, arising from observed changes in the ionising SED. We then constrain the location of the photoionised components in \ngc, study their thermal stability, and investigate their origin and launching mechanism. In Sect. \ref{extended_sect} we study the spatial extent of the X-ray emission in \ngc using \chandra HETG zeroth-order data. We discuss all our findings in Sect. \ref{discussion}, and give concluding remarks in Sect. \ref{conclusions}. 

The spectral analysis and modelling presented here are done using the {\tt SPEX} package \citep{Kaa96} v3.04.00. The spectra shown in this paper are background subtracted and are displayed in the observed frame. We use C-statistics for spectral fitting and give errors at the $1\sigma$ confidence level. We adopt a luminosity distance of 70.55 Mpc in our calculations with the cosmological parameters ${H_{0}=70\ \mathrm{km\ s^{-1}\ Mpc^{-1}}}$, $\Omega_{\Lambda}=0.70$, and $\Omega_{m}=0.30$. We assume proto-solar abundances of \citet{Lod09} in all our computations in this paper.

\section{Observations and data reduction}
\label{data_sect}
%

\subsection{Chandra HETGS data}

All \chandra observations of \ngc have been taken with HETGS \citep{Cani05}. The most recent HETGS observation, hereafter referred to as the 2015 observation, spans over the end of December 2015 and the beginning of January 2016. These new data are presented for the first time in this paper. The archival HETGS observation \citep{Scot05}, hereafter referred to as the 2002 observation, was taken in December 2002 over a period of two days. The logs of the 2002 and 2015 HETGS observations are provided in Table \ref{obs_table}. The total exposure times of these observations are 237~ks (2015) and 147~ks (2002). In all the HETGS observations, the ACIS camera was operated in the Timed Exposure (TE) read mode and the FAINT data mode. The data were reduced using the Chandra Interactive Analysis of Observations ({\tt CIAO}, \citealt{Frus06}) v4.9 software and Calibration Database (CALDB) v4.7.3. The {\tt chandra\_repro} script of {\tt CIAO} and its associated tools were used for the reduction of the data and production of the final grating products (PHA2 spectra, RMF and ARF response matrices). 

The grating spectra and their associated response files were combined using the CIAO {\tt combine\_grating\_spectra} script. The +/- first order spectra of each grating were combined. In addition to producing a HEG and MEG spectrum for each observation, we also produced stacked HEG and MEG spectra containing all the data. We therefore produced three sets of spectra for our spectral modelling: 2002, 2015, and the stacked 2002 and 2015 spectra. The HETGS spectra from the two epochs display similar absorption features and are consistent with each other, thus allowing us to stack the spectra in order to enhance the signal-to-noise ratio. The baseline model derived from the stacked spectra is then applied to the 2002 and 2015 observations to look for variability. In our spectral modelling, HEG and MEG spectra are fitted simultaneously. The fitted spectral range is 2.5--26~\AA\ for MEG, and 1.55--14.5~\AA\ for HEG. Over these energy bands, the HEG/MEG flux ratio is nearly constant at 0.966 in 2002 and 0.956 in 2015. We take into account this instrumental flux difference between HEG and MEG by re-scaling the normalisation of HEG relative to MEG to in our spectral modelling. The low and high energy data outside of these ranges are ignored because of deviations in the HEG/MEG flux ratio caused by the increasing calibration uncertainties of the instruments.

%
\begin{table}[!tbp]
\begin{minipage}[t]{\hsize}
\setlength{\extrarowheight}{3pt}
\caption{Log of \ngc\ \chandra observations. The 2002 observation (147.2 ks exposure time) is split into two parts, while the 2015 observation (236.6 ks exposure time) is split into eight parts.}
\label{obs_table}
\centering
\small
\renewcommand{\footnoterule}{}
\begin{tabular}{c c c c c}
\hline \hline
Obs.		&	Instrument/		&	Start 				& Exposure  \\
ID 		&	Grating			&	Time				& (ks) \\
\hline
2956		&	ACIS-S/HETG		&	2002-12-12 13:37	& 78.58 \\
3147		&	ACIS-S/HETG		&	2002-12-13 12:10	& 68.63 \\
\hline
18622	&	ACIS-S/HETG		&	2015-12-27 05:46	& 43.31 \\
18733	&	ACIS-S/HETG		&	2015-12-29 00:47	& 17.48 \\
18734	&	ACIS-S/HETG		&	2015-12-29 20:30	& 29.53 \\
18735	&	ACIS-S/HETG		&	2015-12-31 14:57	& 21.15 \\
18736	&	ACIS-S/HETG		&	2016-01-01 12:56	& 31.01 \\
18737	&	ACIS-S/HETG		&	2016-01-03 03:05	& 31.28 \\
18623	&	ACIS-S/HETG		&	2016-01-04 21:44	& 36.03 \\
18738	&	ACIS-S/HETG		&	2016-01-06 19:11	& 26.78 \\
\hline
\end{tabular}
\end{minipage}

\end{table}

\subsection{Swift UVOT and HST COS data}

In order to determine the SED of \ngc, we made use of optical/UV data, taken contemporaneously with the HETGS observations: \swift UVOT \citep{Romi05} and \hst COS \citep{Green12} 2015 data, as well as HST STIS \citep{Wood98} 2002 data. The \swift UVOT data from Image-mode operations were taken with the six primary photometric filters of V, B, U, UVW1, UVM2 and UVW2. The $\mathtt{uvotsource}$ tool was used to perform aperture photometry using a circular aperture diameter of 10$\arcsec$. The standard instrumental corrections and calibrations according to \citet{Poo08} were applied. Any data with bad \swift tracking were discarded. For the purpose of spectral fitting with {\tt SPEX}, the count rate and the corresponding response matrix file for each filter were created.

The \hst COS observations through the Primary Science Aperture were taken with gratings G130M and G160M, covering the far-UV spectral range from 1132 to 1801 \AA. In addition to the routine processing of data with the calibration pipelines of STScI, further enhanced calibrations (as described in \citealt{Kris11}) were applied to produce the best-quality COS spectra possible. They include refined flux calibrations that take into account up-to-date adjustments to the time-dependent sensitivity of COS, improved flat-field corrections and wavelength calibration, and an optimal method for combining exposures comprising a single visit. Further details on the spectral analysis of the \ngc COS spectra will be given in \citet{Arav17}. In this paper, for the purpose of modelling the UV continuum as part of our broadband continuum modelling, we extracted the \hst COS (2015) and STIS (2002, \citealt{Scot05}) fluxes from three narrow spectral bandpasses, which are free of emission and absorption lines: 1162--1178, 1478--1498, 1720--1730 \AA. 

\section{Spectral energy distribution determination}
\label{sed_sect}

\begin{figure}[!tbp]
\centering
\resizebox{\hsize}{!}{\includegraphics[angle=0]{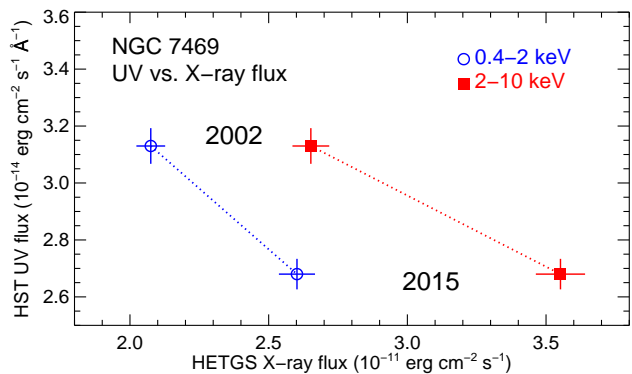}}
\caption{Comparison of the UV (HST) and X-ray (\chandra) fluxes of \ngc in the 2002 and 2015 epochs. The HST flux at $1170~\AA$ is plotted versus the HETGS flux over 0.4--2 keV ({\it shown in blue}) and 2--10 keV ({\it shown in red}). The HST data are from STIS (2002) and COS (2015) observations. The displayed fluxes are the observed fluxes without any modification for reddening or absorption.}
\label{flux_fig}
\end{figure}

The ionisation and thermal equilibrium in a photoionised plasma is strongly dependent on the SED of the ionising source. The 2002 and 2015 UV and X-ray fluxes indicate a change in the shape of the SED. As shown in Fig. \ref{flux_fig}, the UV flux is lower in 2015 than in 2002, but both the soft and hard X-ray fluxes are higher in 2015 than in 2002. Therefore, we determine the SED of \ngc for both the 2002 and 2015 epochs. We describe in the following how the intrinsic continuum and the contaminating non-intrinsic emission and absorption components in our line of sight to \ngc were modelled.

\subsection{Primary optical-UV-X-ray continuum of the AGN}
\label{continuum_sect}

The observed primary hard X-ray continuum is modelled with a power-law component ({\tt pow}) to represent Compton up-scattering of the disk photons in an optically-thin, hot, corona in \ngc \citep{Petr04}. We apply a high-energy exponential cut-off to the power-law at 170~keV, which is based on the analysis of the 2015 \nustar observations \citep{Midd17}. A low-energy exponential cut-off was also applied to the power-law continuum to prevent exceeding the energy of the seed disk photons. We find that while the power-law flux is higher in 2015 than in 2002, the photon index $\Gamma$ of the power-law is the same in both epochs (${\Gamma = 1.91 \pm 0.01}$).

\begin{figure}[!tbp]
\centering
\resizebox{\hsize}{!}{\includegraphics[angle=0]{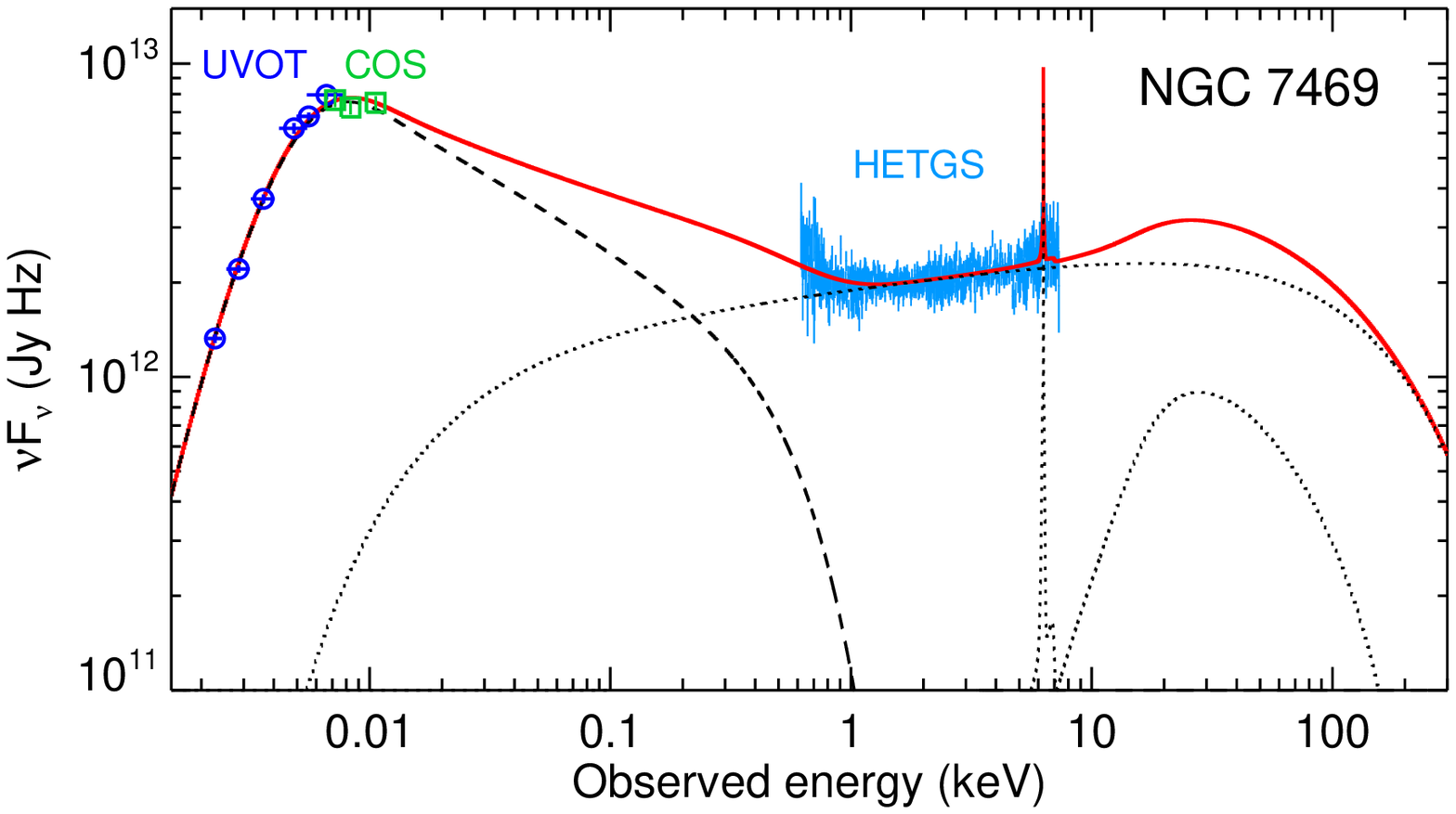}}\vspace{-0.3cm}
\resizebox{\hsize}{!}{\includegraphics[angle=0]{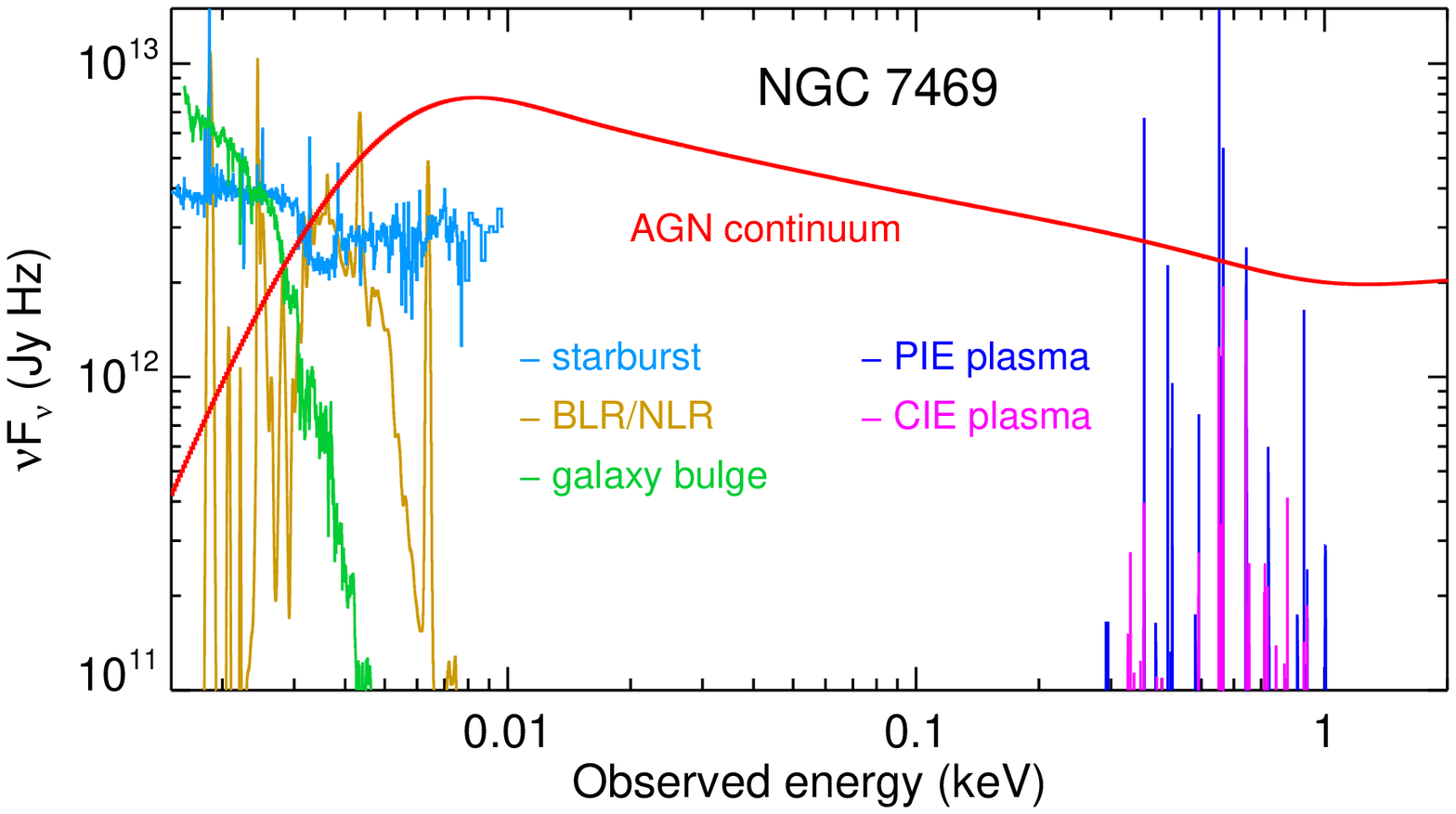}}\vspace{-0.3cm}
\resizebox{\hsize}{!}{\includegraphics[angle=0]{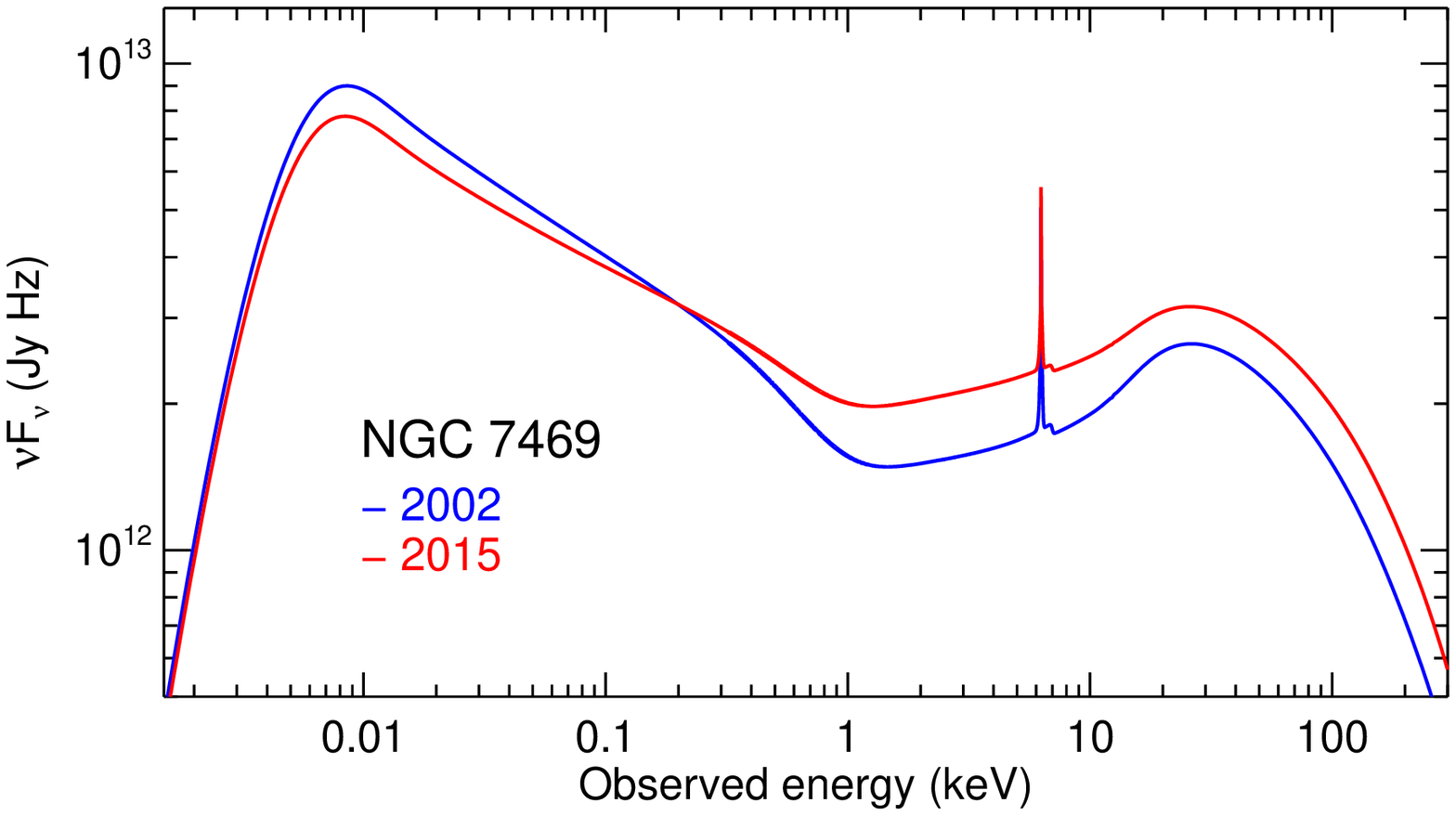}}
\caption{Spectral energy distributions of \ngc from optical to hard X-rays. {\it Top panel}: the best-fit continuum model, derived in Sect. \ref{sed_sect}, fitted to the 2015 \chandra HETGS, \hst COS and \swift UVOT data. The displayed data are corrected for the emission and absorption effects described in Sect. \ref{sed_sect}, required for determining the underlying AGN continuum. The individual components of the continuum model are displayed: a warm Comptonisation component ({\tt comt}), shown in dashed black line, a power-law ({\tt pow}) and a reflection ({\tt refl}) component, shown in dotted black lines. {\it Middle panel}: the 2015 AGN continuum model (shown in red), and the individual optical/UV and soft X-ray emission components that were modelled for uncovering the underlying continuum. {\it Bottom panel}: comparison of the SED continuum models for the 2002 and 2015 epochs, which were used in our photoionisation modelling. The 1--1000 Ryd luminosity is ${1.37 \times 10^{44}}$~\ergs in 2002 and ${1.42 \times 10^{44}}$~\ergs in 2015. The bolometric luminosity is ${3.4 \times 10^{44}}$~erg~s$^{-1}$ in 2002 and ${3.8 \times 10^{44}}$~erg~s$^{-1}$ in 2015.}
\label{sed_fig}
\end{figure}

In addition to the power-law, the soft X-ray continuum of \ngc shows the presence of a ``soft X-ray excess'' component \citep{Barr86,Blu03}, with similar characteristics to those found in the archetypical Seyfert-1 AGN \object{NGC 5548} \citep{Meh15a}. To model the soft excess in \ngc, we use the broadband model derived in \citet{Meh15a} for NGC~5548, in which the soft excess is modelled by warm Comptonisation (see e.g. \citealt{Mag98, Meh11, Done12, Petr13, Petr17}). In this explanation of the soft excess, the disk seed photons are up-scattered in an optically thick, warm, corona to produce the soft X-ray excess. In this scenario the strength of the soft excess component is correlated to that of the optical/UV emission from the accretion disk. For more details on the theory of soft excess emission from the disk, see \citet{Done12,Petr13,Petr17}. The {\tt comt} model in \spex produces a thermal optical/UV disk component modified by warm Comptonisation, so that its high-energy tail fits the soft X-ray excess. The {\tt comt} model is also useful for photoionisation modelling, because it determines the strength of the unobservable EUV emission, which strongly influences the ionisation balance of photoionised gas. In recent years, multi-wavelength studies have found warm Comptonisation to be a viable explanation for the soft excess in Seyfert-1 AGN (e.g. most recently in \citealt{Petr17,Porq17}). We find that the {\tt comt} model provides a good fit to the optical-UV-X-ray data of \ngc (Fig. \ref{sed_fig}, top panel). The fitted parameters of the {\tt comt} model are its normalisation, seed photons temperature $T_{\rm seed}$, electron temperature $T_{\rm e}$, and optical depth $\tau$ of the up-scattering plasma. The best-fit parameters of the power-law component and the warm Comptonisation component (disk + soft X-ray excess) are given in Table \ref{refl_table}.

The optical/UV continuum, modelled with {\tt comt}, has an intrinsic luminosity of ${6.3 \pm 0.3 \times 10^{43}}$~\ergs in 2002, and ${5.5 \pm 0.3 \times 10^{43}}$~\ergs in 2015 over 1000--7000 \AA. The luminosity of the soft X-ray excess (0.2--2 keV) follows a similar trend, dropping from ${10.0 \pm 0.5 \times 10^{42}}$~\ergs in 2002 to ${8.2 \pm 0.4 \times 10^{42}}$~\ergs in 2015. On the other hand, the luminosity of the X-ray power-law over 0.2--10 keV shows the opposite behaviour by increasing from ${3.5 \pm 0.1 \times 10^{43}}$~\ergs in 2002 to ${4.7 \pm 0.1 \times 10^{43}}$~\ergs in 2015. Such trends in the variability of the continuum components have been seen in the multiwavelength studies of the soft excess. For example, in the study of \mrk using extensive optical/UV and X-ray monitoring data, \citet{Meh11} found that the variability of the soft X-ray excess is strongly correlated with that of the optical/UV emission from the disk. On the other hand, the variability of the X-ray power-law component was found to be uncorrelated with those of the soft X-ray excess and the optical/UV emission. This supports the two-component model for the primary X-ray continuum used here.

In addition to the primary X-ray continuum, a reprocessed component produced by X-ray reflection, is also taken into account, which is described in the following.

%
 
%

\subsection{Reflected X-ray continuum}
\label{refl_sect}

\begin{figure}[!tbp]
\centering
\vspace{-0.7cm}
\resizebox{1.1\hsize}{!}{\hspace{-1.3cm}\includegraphics[angle=270]{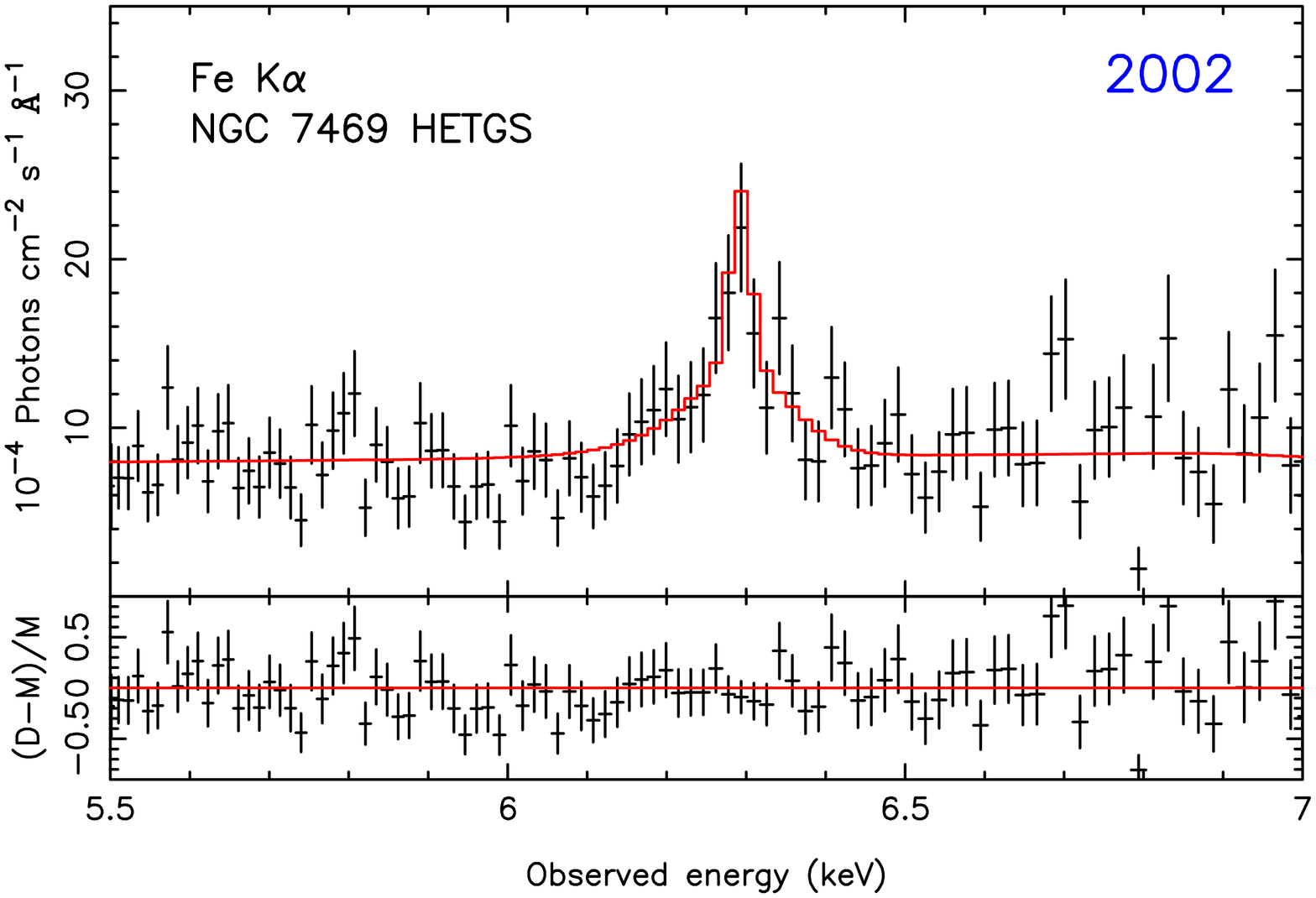}}\vspace{-1.4cm}
\resizebox{1.1\hsize}{!}{\hspace{-1.3cm}\includegraphics[angle=270]{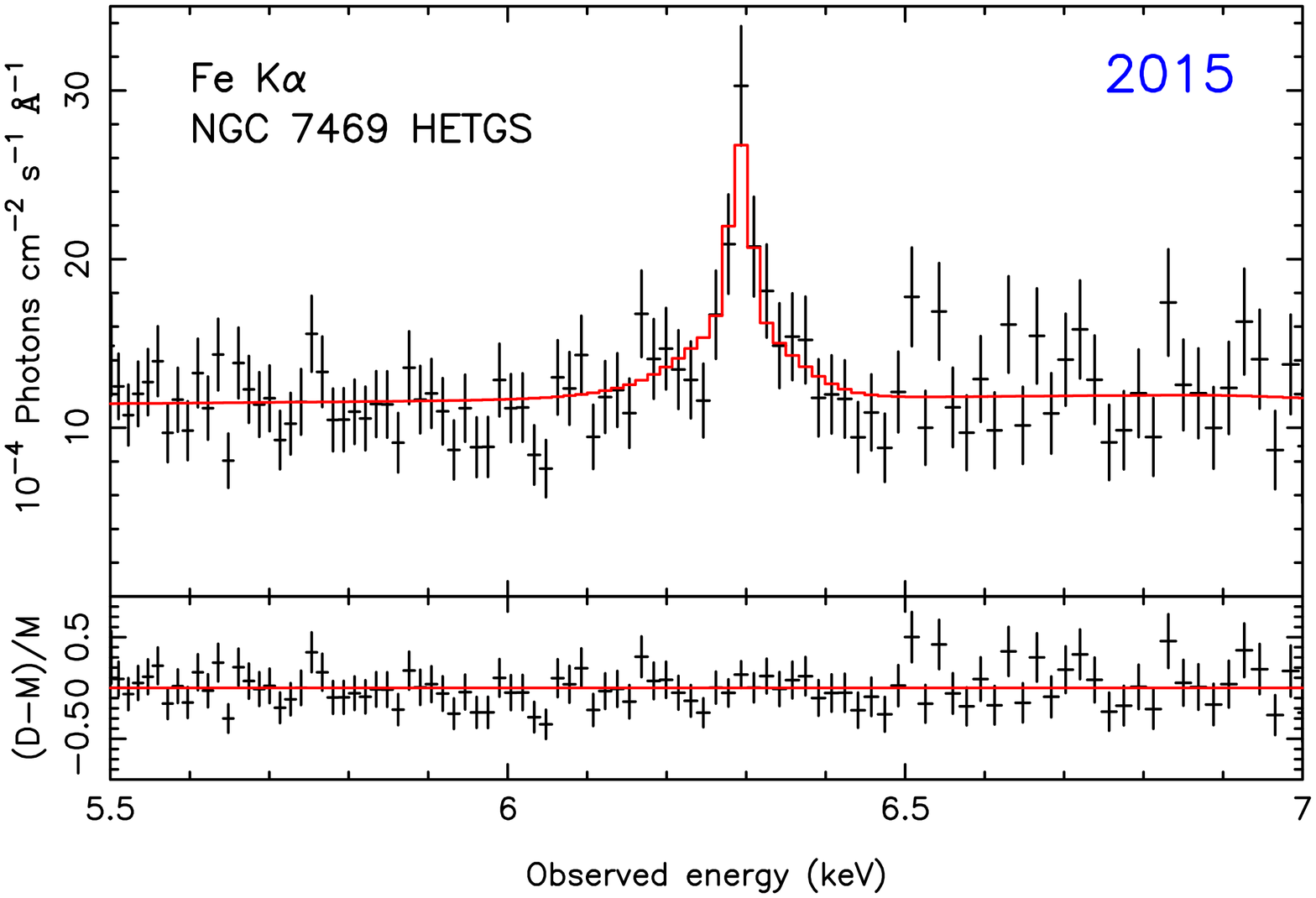}}
\vspace{-0.4cm}
\caption{\FeKa line of \ngc and its best-fit model from the \chandra HETGS (HEG) 2002 ({\it upper panel}) and 2015 ({\it lower panel}) spectra. The applied model is described in Sect. \ref{refl_sect}, with its parameters given in Table \ref{refl_table}. The fit residuals are shown in the bottom panel for each observation.}
\label{FeKa_fig}
\end{figure}

The HETGS (HEG) spectra of \ngc show a clear presence of \FeKa line emission (Fig. \ref{FeKa_fig}). We applied an X-ray reflection component ({\tt refl}), which reprocesses an incident power-law continuum to produce the \FeKa line and the Compton hump at hard X-rays. The {\tt refl} model in \spex computes the \FeKa line according to \citet{Zyck94}, and the Compton-reflected continuum according to \citet{Magd95}, as described in \citet{Zyck99}. The photon index $\Gamma$ of the incident power-law was set to that of the observed primary continuum, which is $\Gamma = 1.91\pm 0.01$ for both the 2002 and 2015 epochs. The exponential high-energy cut-off of this incident power-law is also set to that of the observed primary power-law component at 170~keV, based on the \nustar measurements \citep{Midd17}. In our modelling we fitted the normalisation of the incident power-law continuum and the reflection scale, $s$. The ionisation parameter of {\tt refl} is set to zero to produce a cold reflection component with all abundances kept at their solar values. The {\tt refl} model was convolved with a Gaussian velocity broadening model to fit the width $\sigma_{v}$ of the \FeKa line, which is about $2700 \pm 800$~\kms in both epochs. We do not require a relativistic line profile to fit the \FeKa line (Fig. \ref{FeKa_fig}).

%

Our analysis of the \FeKa line shows that the line profile is composed of two components: the main broad component ($2700$~\kms), fitted with the {\tt refl} model, and a narrow unresolved core, which we fitted with a delta line model ({\tt delt}). We fitted the 2002 and 2015 HETGS (HEG) spectra with the same {\tt refl} and {\tt delt} model, and allowed the normalisation of the incident power-law continuum for {\tt refl} to vary between the observations. We find this alone provides a good fit to the data for both epochs (Fig. \ref{FeKa_fig}). Our model parameters for the \FeKa line are provided in Table \ref{refl_table}. The 2--10 keV intrinsic flux of the incident power-law is ${6.0 \times 10^{-11}}$ \ergflux in 2002 and ${5.1 \times 10^{-11}}$ \ergflux in 2015. The derived normalisations of the incident power-law for the 2002 and 2015 observations are within 1-$\sigma$ errors; therefore, X-ray reflection is consistent with being unchanged in both epochs. We note that the associated Compton hump produced from our HETGS modelling of the \FeKa line using the {\tt refl} model, is consistent with that obtained from modelling the 2015 \nustar data \citep{Midd17}. The 10--100 keV flux of the Compton hump according to our model is ${1.7 \times 10^{-11}}$ \ergflux in 2002 and ${1.4 \times 10^{-11}}$ \ergflux in 2015.

\subsection{Galactic interstellar absorption and reddening}

The continuum and line absorption by the interstellar medium (ISM) of the Galaxy are included in our modelling of the X-ray spectra by applying the {\tt hot} model in \spex. This model calculates the transmission of a plasma in collisional ionisation equilibrium (CIE) at a given temperature, which in the case of cold ISM is set to 0.5~eV. The total Galactic \NH column density is fixed to $N_{\mathrm{H}}={5.49\times 10^{20}\ \mathrm{cm}^{-2}}$, which is the sum of the atomic \citep{Wakk11} and molecular \citep{Wakk06} \NH components of the ISM in our line of sight to \ngc.

To correct the optical/UV data (\swift UVOT and \hst COS) for Galactic interstellar dust reddening, we applied the {\tt ebv} model in \spex. The model incorporates the reddening curve of \citet{Car89}, including the update for near-UV given by \citet{ODo94}. We initially set the colour ${E(B-V) = 0.061}$~mag \citep{Schl11}, and the scalar specifying the ratio of total to selective extinction $R_V \equiv A_V/E(B-V)$ was fixed to 3.1. In order to take into account additional reddening by the host galaxy of the AGN, we allowed the $E(B-V)$ parameter to be fitted in our modelling, which increased slightly to ${0.082 \pm 0.003}$ for both epochs.

\subsection{Optical/UV emission from the BLR and the NLR}
\label{BLR_sect}

Apart from the optical/UV continuum, the photometric filters of \swift UVOT contain emission from the broad-line region (BLR) and the narrow-line region (NLR) of the AGN. Therefore, in order to correct for this contamination, we applied the emission model derived in \citet{Meh15a} for NGC~5548 as a template model to the optical/UV data of \ngc. The model takes into account the Balmer continuum, the \ion{Fe}{ii} feature, and the emission lines from the BLR and NLR. We freed the normalisation of this component in our modelling of the \ngc UVOT data. The predicted H$\beta$ flux from our fit is ${5.9 \times 10^{-13}}$ \ergflux in 2015, while the optical continuum flux level is at ${1.4 \times 10^{-14}}$ \ergcm over the 5180--5200~\AA\ band. These measurements are consistent with the trend of H$\beta$ flux variability in \ngc as reported by the long-term monitoring study of \citet{Shap17}.
 
%
\begin{table}[!tbp]
\begin{minipage}[t]{\hsize}
\setlength{\extrarowheight}{3pt}
\caption{Best-fit parameters of our model for the broadband continuum and the \FeKa line of \ngc, obtained from the 2002 and 2015 \chandra HETGS, \hst COS and \swift UVOT observations.}
\label{refl_table}
\centering
\small
\renewcommand{\footnoterule}{}
\begin{tabular}{l | c}
\hline \hline
Parameter						& Value					\\
\hline
\multicolumn{2}{c}{Primary power-law component ({\tt pow}):} 						\\
Normalisation			& ${5.3 \pm 0.1}$~(2002), ${7.1 \pm 0.1}$~(2015) \\
Photon index $\Gamma$		& ${1.91 \pm 0.01}$			\\
\hline
\multicolumn{2}{c}{Disk + soft X-ray excess component ({\tt comt}):} 						\\
Normalisation	&		${6.5 \pm 0.3}$~(2002), ${5.7 \pm 0.3}$~(2015) \\
$T_{\rm seed}$ (eV) &	${1.4 \pm 0.1}$ \\
$T_{\rm e}$ (keV) &		${0.14 \pm 0.02}$ \\
Optical depth $\tau$ &	${21 \pm 4}$ \\
\hline
\multicolumn{2}{c}{Resolved \FeKa component ({\tt refl}):} 						\\
Incident power-law Norm.			& ${12 \pm 4}$~(2002), ${10 \pm 4}$~(2015) \\
Incident power-law $\Gamma$		& $1.91$ (f)		 			\\
Reflection scale	$s$			& $0.30 \pm 0.10$					\\
$\sigma_v$ (\kms)				& $2700 \pm 800$					\\
\hline
\multicolumn{2}{c}{Unresolved \FeKa component ({\tt delt}):}		 						\\
Flux (${10^{-13}}$~\ergflux)		& $1.2 \pm 0.3$			\\
$E_0$ (keV)					& $6.396 \pm 0.005$			\\
$\sigma_v$ (\kms)				& $< 460$ 				\\
\hline
\end{tabular}
\end{minipage}
\tablefoot{
The power-law normalisation of {\tt pow} and {\tt refl} components is in units of $10^{51}$ photons~s$^{-1}$~keV$^{-1}$ at 1 keV. The normalisation of the warm Comptonisation component ({\tt comt}) is in units of $10^{55}$ photons~s$^{-1}$~keV$^{-1}$. The exponential high-energy cut-off of the power-law for both {\tt pow} and {\tt refl} is fixed to 170~keV (based on \nustar measurements). The photon index $\Gamma$ of the incident power-law for the reflection component ({\tt refl}) is set to the $\Gamma$ of the observed primary power-law continuum ({\tt pow}). The parameters that have been coupled between the two epochs in our modelling have a single value in the table.
}
\end{table}
 
\subsection{Emission from the nuclear starburst ring and the galactic bulge in \ngc}
\label{bulge_sect}

\begin{figure}[!tbp]
\centering
\resizebox{0.8\hsize}{!}{\includegraphics[angle=0]{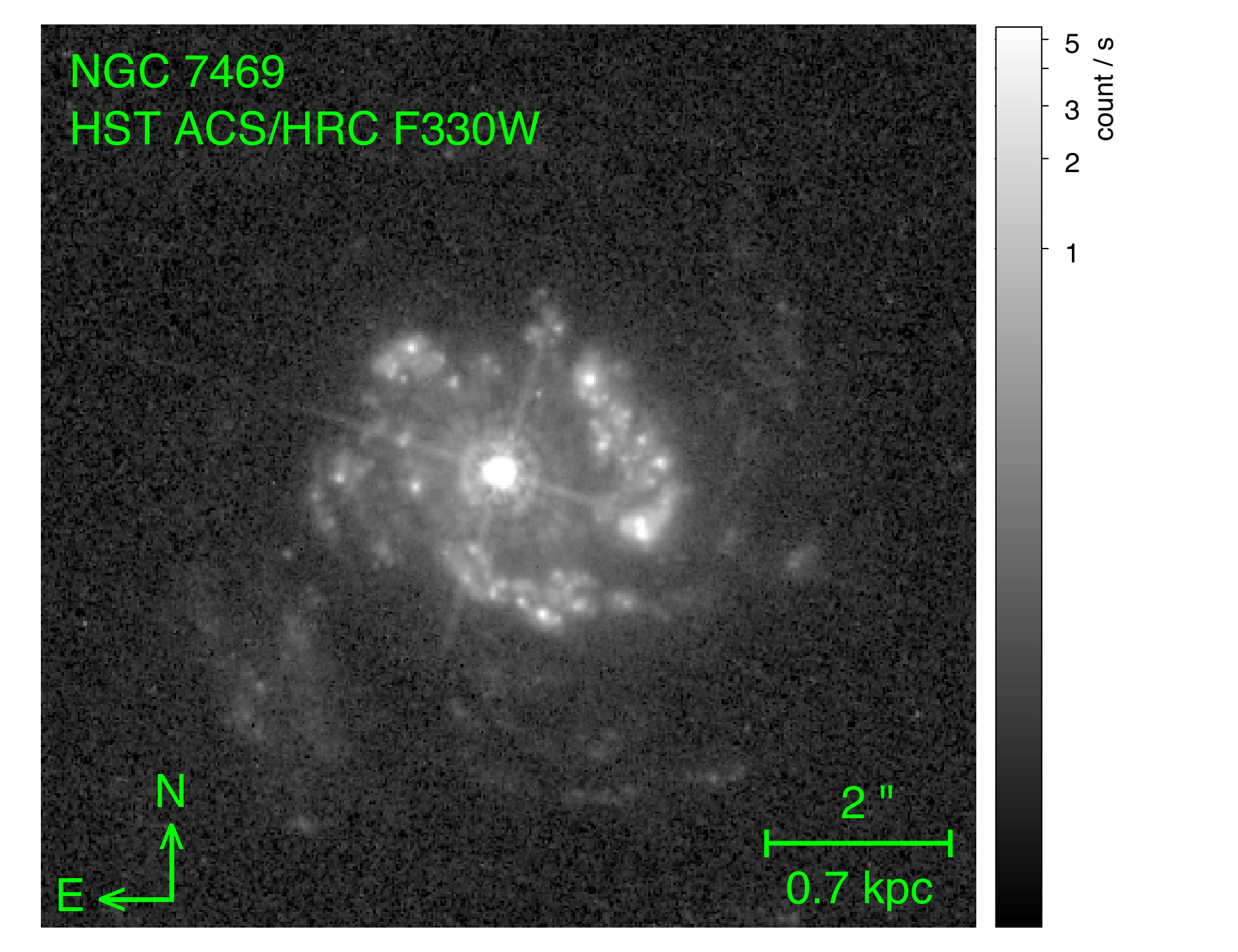}\hspace{-2.8cm}}
\caption{Nuclear region of \ngc as observed by HST. The image is obtained from an observation taken with the ACS/HRC F330W filter in 2002, and is displayed with a logarithmic intensity scale. The inner starburst ring is visible at about 1--3$\arcsec$ from the centre of the AGN. The outer starburst ring (not displayed) is at 8--10$\arcsec$ from the centre. For reference, the circular aperture of our \swift UVOT observations has a radius of 5$\arcsec$. We find an extended soft X-ray emission component by \chandra at 1.5--12$\arcsec$ from the centre (Fig. \ref{radial_fig}).}
\label{starburst_fig}
\end{figure}

The nuclear region of \ngc exhibits intense star formation with a dual starburst ring structure surrounding the nucleus (e.g. \citealt{Maud94}). The inner and outer starburst rings are at about 1--3$\arcsec$ and 8--10$\arcsec$ radii from the centre of the AGN. We take into account the contribution of stellar emission from the host galaxy of \ngc in our SED modelling. The circular aperture of UVOT (5$\arcsec$ radius) takes the inner nuclear starburst ring and part of the galactic bulge in \ngc. A HST image of the inner starburst ring is shown in Fig. \ref{starburst_fig}, which we extracted from a 1.2 ks observation taken with the ACS/HRC F330W filter on 20 November 2002.

\citet{Ben13} have determined the optical flux of the host galaxy components for a number of AGN (including \ngc) using modelling of HST images. The galaxy bulge and the nuclear starburst ring flux were re-calculated by M. Bentz for our UVOT aperture size. From modelling of HST images taken with the F547M filter, the optical flux at $5447~\AA$ is determined to be ${8.33 \times 10^{-15}}$ \ergcm for the galaxy bulge, and ${2.97 \times 10^{-15}}$ \ergcm for the inner starburst ring (Bentz, priv. comm.). The uncertainty in these flux measurements is about $10\%$. We then adopted appropriate spectral model templates for these host galaxy components, and normalised them to the above monochromatic flux measurements for \ngc. This way the contribution of the host galaxy components to the observed flux in the bandpasses of the UVOT filters is calculated. We incorporated the galaxy bulge and the SB3 starburst template models of \citet{Kin96} in our SED modelling. In Fig. \ref{sed_fig} (middle panel), the optical/UV spectra of the bulge and the starburst ring in \ngc are displayed. We can see that in the energy range of the UVOT filters, emission from the starburst and the galactic bulge dominate over the AGN continuum. Therefore, proper modelling of these non-intrinsic components, like performed here, are important for deriving the shape of the underlying AGN optical/UV continuum.  

In \citet{Beh17} we reported the detection of coronal X-ray emission lines of \ion{Fe}{xvii} in the stacked 640~ks RGS spectrum of \ngc, at the observed wavelengths of 15.25 \AA\ and 17.38 \AA. These lines likely originate from the nuclear starburst. Thus, in our modelling of the HETGS spectra, we included a \cie component in \spex to model the emission from such a CIE plasma. The emission measure (EM), temperature, and outflow velocity of this \cie component were set to those found by \citet{Beh17}: EM~${= 4 \times 10^{63}}$~cm$^{-3}$, ${T=0.35}$~keV, and ${v_{\rm out} = -250}$~\kms. We refitted these parameters in our modelling of the HETGS data, which we find to remain unchanged within errors. At the temperature of the \cie model, the concentration of the Fe ions peaks at \ion{Fe}{xvii}. Later in Sect. \ref{extended_sect}, we analyse the \chandra ACIS/HETG zeroth-order image and spectrum to determine the spatial extent of X-ray emission in \ngc. We indeed find a diffuse soft X-ray component at 1.5--12$\arcsec$ radii from the central source, which our modelling shows to be consistent with the aforementioned CIE component from the nuclear starburst in \ngc.

%

\section{Modelling of the AGN wind in \ngc}
\label{model_sect}
%

\subsection{Photoionisation modelling}
\label{photoionisation_sect}

For photoionisation modelling and spectral fitting, we use the \pion model in \spex, which is a self-consistent model that calculates the thermal/ionisation balance together with the spectrum of a plasma in photoionisation equilibrium (PIE). The \pion model uses the SED (Sect. \ref{sed_sect}) from the continuum model components set in \spex. During spectral fitting, as the continuum varies, the thermal/ionisation balance and the spectrum of the plasma are re-calculated at each stage. This means while using realistic broadband continuum components to fit the data, the photoionisation is calculated accordingly by the \pion model. This is useful as the derived shape of the SED can significantly influence the structure and thermal stability of the AGN winds (see e.g. \citealt{Chak12,Meh16b}). For more details about the \pion model and its comparison with other photoionisation codes see \citet{Meh16b}. In our computations of the photoionisation equilibrium and the X-ray spectrum, the elemental abundances are fixed to the proto-solar values of \citet{Lod09}.

\subsection{HETGS spectroscopy of the wind}
\label{spectroscopy_sect}

\begin{figure}[!tbp]
\centering
\resizebox{0.8\hsize}{!}{\includegraphics[angle=0]{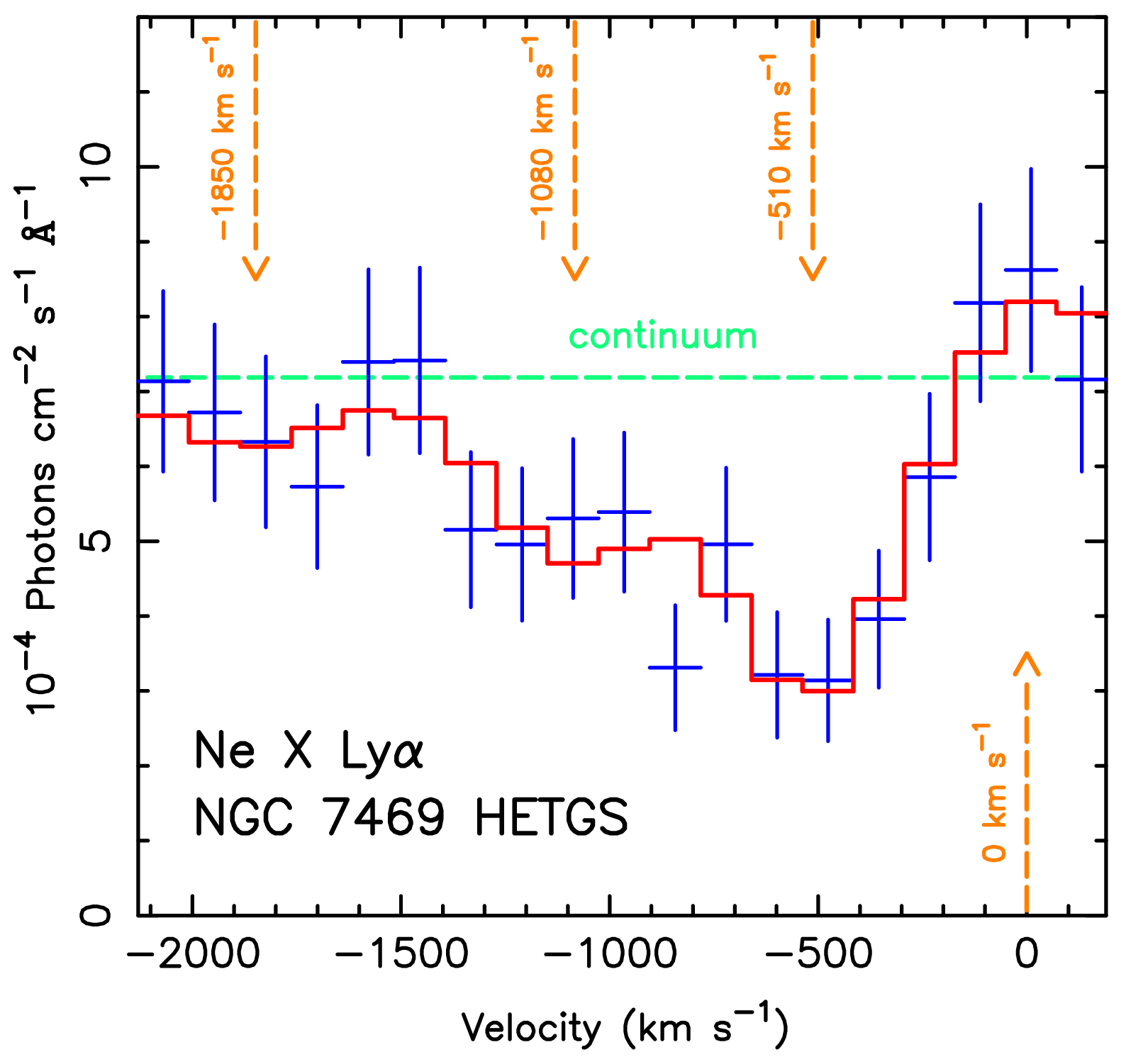}}
\caption{Absorption profile of the \ion{Ne}{x} Ly$\alpha$ line (12.135 \AA) in \ngc as seen in the HETGS (MEG) spectrum. The contribution to absorption from three outflowing components are indicated. An emission component is also present at rest velocity.}
\label{NeX_fig}
\end{figure}

\begin{figure*}[!tbp]
\vspace{-1cm}
\hspace{-1.3cm}\resizebox{1.15\hsize}{!}{\includegraphics[angle=270]{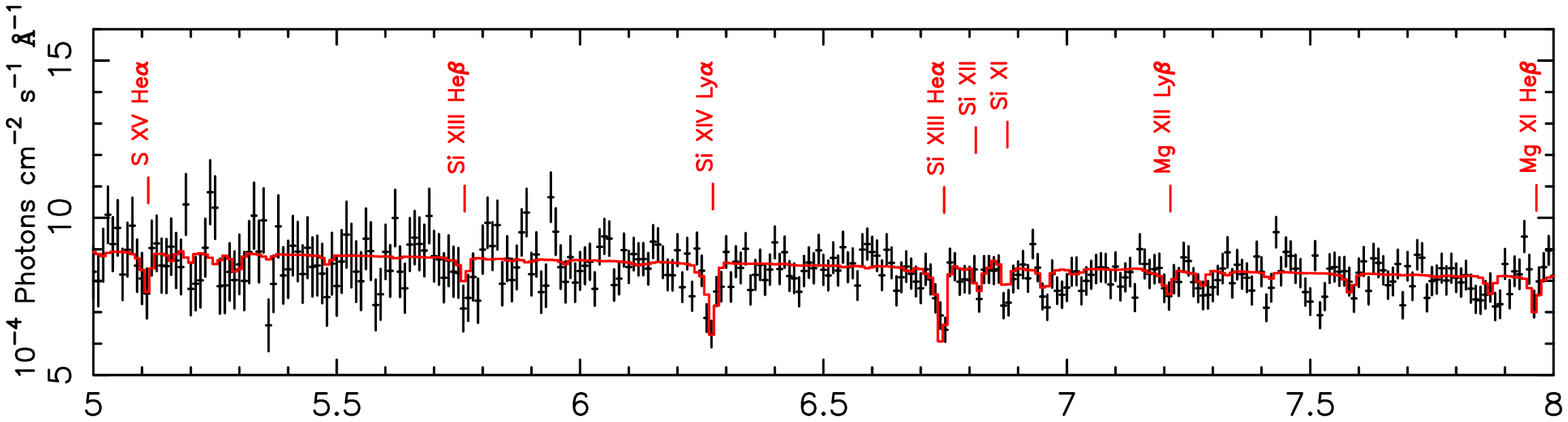}}\vspace{-10.4cm}

\hspace{-1.3cm}\resizebox{1.15\hsize}{!}{\includegraphics[angle=270]{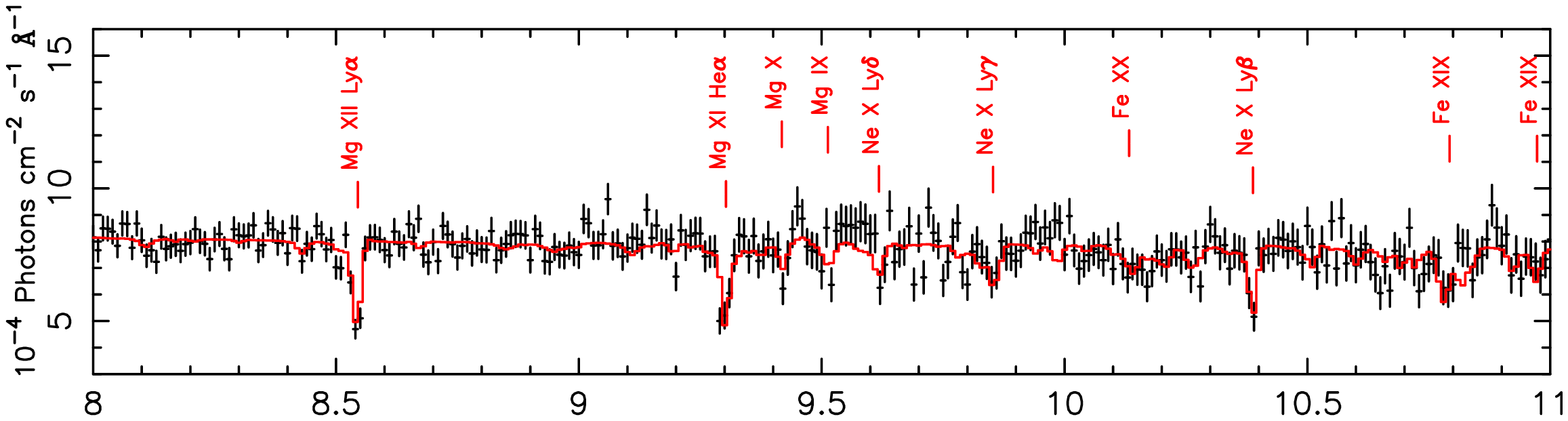}}\vspace{-10.4cm}

\hspace{-1.3cm}\resizebox{1.15\hsize}{!}{\includegraphics[angle=270]{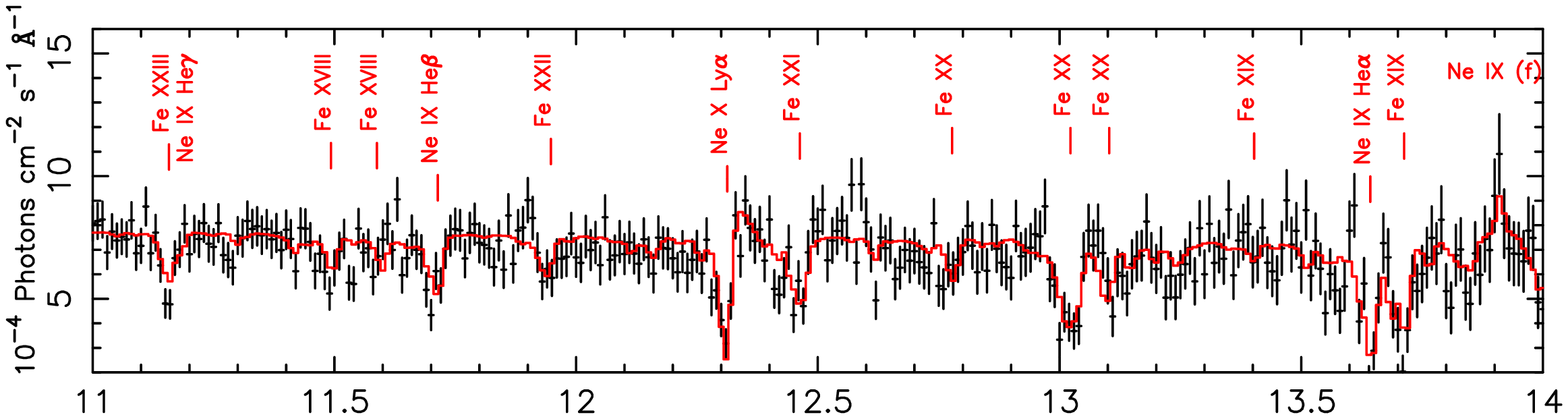}}\vspace{-10.4cm}

\hspace{-1.3cm}\resizebox{1.15\hsize}{!}{\includegraphics[angle=270]{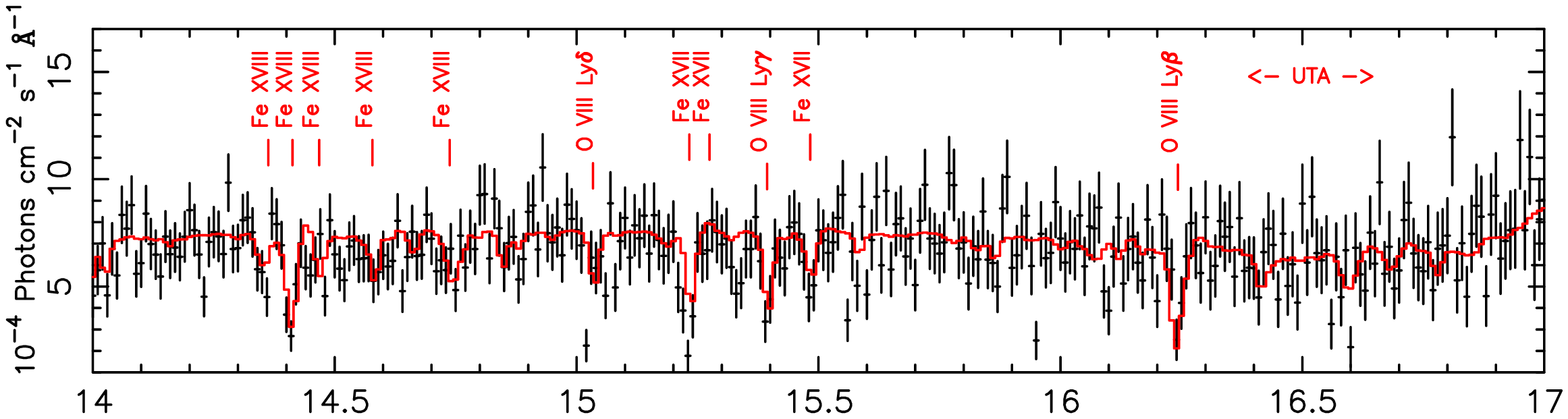}}\vspace{-10.4cm}

\hspace{-1.3cm}\resizebox{1.15\hsize}{!}{\includegraphics[angle=270]{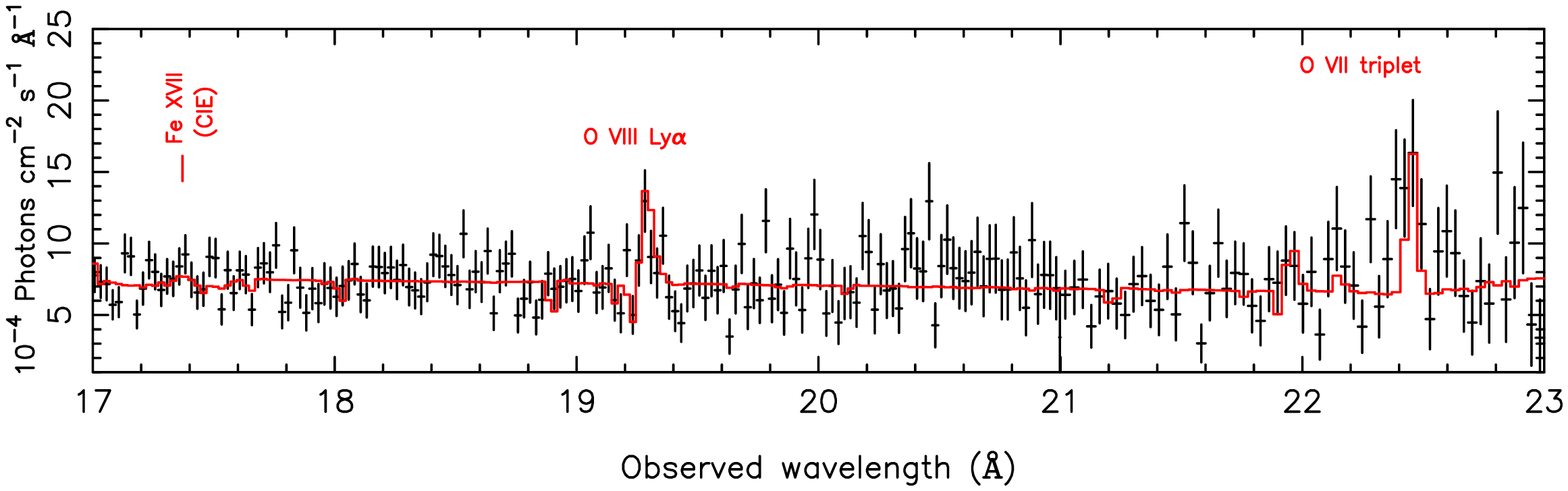}}\vspace{-8.1cm}

\caption{\chandra HETGS (MEG) stacked spectrum of \ngc and its best-fit model (Table \ref{pion_table}).}
\label{meg_fig}
\end{figure*}

We first obtained a best-fit model to the stacked (2002 + 2015) HETGS spectrum of \ngc, and then used this model as a starting point to fit the individual 2002 and 2015 HETGS spectra. The HETGS spectrum of \ngc exhibits a series of clear spectral lines (see Figs. \ref{NeX_fig} and \ref{meg_fig}). To properly fit all the absorption lines in the HETGS spectrum from various ionic species we require four \pion components with a different ionisation parameter $\xi$ \citep{Tar69,Kro81}. This parameter is defined as ${\xi \equiv L / n_{\rm H}\, r^2}$, where $L$ is the luminosity of the ionising source over the 1--1000 Ryd band (13.6 eV to 13.6 keV) in \ergs, $n_{\rm H}$ the hydrogen density in cm$^{-3}$, and $r$ the distance between the photoionised gas and the ionising source in cm. The lowest ionisation component (Comp. A) according to our modelling produces absorption from the M-shell Fe ions to form a shallow Unresolved Transition Array (UTA, \citealt{Beh01}) at about 16--17 \AA. At $\xi$ of Comp. A, the ionic density of the Fe ions producing the UTA peaks at \ion{Fe}{XI}. Moreover, Comp. A is responsible for producing lines from the Be-like and Li-like \ion{Si}{xi}, \ion{Mg}{ix}, and \ion{Mg}{x} ions, as well as the He-like \ion{Ne}{ix}. The next ionisation component (Comp. B) primarily produces lines from the He-like \ion{Ne}{ix} and \ion{Mg}{xi} ions, as well as the H-like \ion{O}{viii} and \ion{Ne}{x}. This component is also responsible for lines of \ion{Fe}{xvii} to \ion{Fe}{xix}. The next ionisation component (Comp. C) mainly produces lines from the H-like \ion{Ne}{x}, \ion{Mg}{xii}, and \ion{Si}{xiv} ions, together with lines from \ion{Fe}{xx} and \ion{Fe}{xxi}. Finally, the highest ionisation component (Comp. D) mostly produces lines from \ion{Fe}{xxi}, \ion{Fe}{xxii}, and \ion{Fe}{xxiii} in the HETGS spectrum. We do not detect line absorption by the more ionised H-like and He-like Fe ions in the Fe-K band (Fig. \ref{FeKa_fig}).

%

%
\begin{figure}[!tbp]
\vspace{-0.5cm}
\centering
\resizebox{0.95\hsize}{!}{\includegraphics[angle=0]{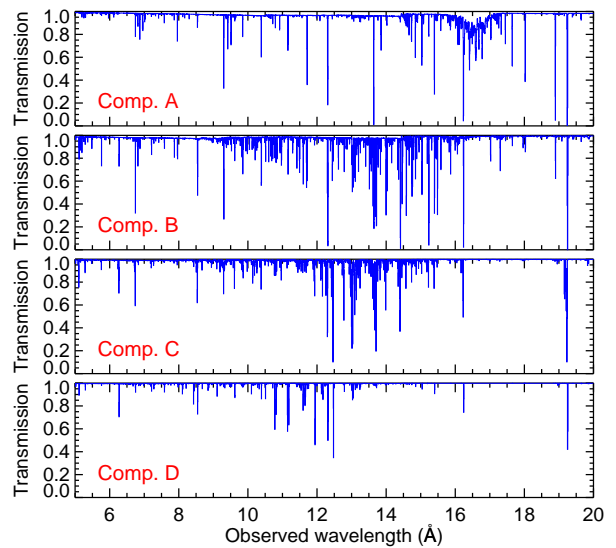}}
\caption{Model transmission X-ray spectrum of the four \pion absorption components of the AGN wind in \ngc, where Comp. A is the lowest ionisation component, and Comp. D the highest. The model is derived from the stacked HETGS spectrum as described in Sect. \ref{model_sect}. The parameters of each component are given in Table \ref{pion_table}.}
\label{trans_fig}
\end{figure}

%

%
\begin{table*}[!tbp]
\begin{minipage}[t]{\hsize}
\setlength{\extrarowheight}{3pt}
\caption{Best-fit parameters of the \pion photoionisation model components fitted to the stacked \chandra HETGS spectrum of \ngc as described in Sect. \ref{model_sect}. Ionisation components A to D are the absorption components of the AGN wind, and E and F the emission components.}
\label{pion_table}
\centering
\small
\renewcommand{\footnoterule}{}
\begin{tabular}{c | c c c c | c | c}
\hline \hline
Ionisation		& $\log~\xi$		 	& \NH 			& $v_{\rm out}$  								& $\Omega\,/\,4 \pi$ 			&	EM					& $\Delta$\,C-stat \\
Component	& (erg~cm~s$^{-1}$)	 	& ($10^{21}$~\cm)	& (\kms)										&						&	($10^{64}$~cm$^{-3}$)	& 		\\
\hline
A			& $1.90 \pm 0.05$		& $0.7 \pm 0.1$	& $-540 \pm 40$								& -						& -						& 303	\\
B			& $2.40 \pm 0.03$		& $1.8 \pm 0.2$	& $-460 \pm 10$								& -						& -						& 584	\\
C			& $2.79 \pm 0.03$		& $1.7 \pm 0.2$	& $-700 \pm 30$, $-1080 \pm 60$, $-1830 \pm 100$		& -						& -						& 291	\\
D			& $3.28 \pm 0.04$	 	& $2.3 \pm 0.7$	& $-370 \pm 30$								& -						& -						& 49		\\
\hline
E			& $0.9 \pm 0.4$	 	& $26 \pm 12$		& $-220 \pm 50$								& $0.006 \pm 0.002$			& $5.7$					& 37		\\
F			& $2.2 \pm 0.1$	 	& $120 \pm 45$	& $0$ (f)										& $0.005 \pm 0.002$			& $1.0$					& 79		\\
\hline
\multicolumn{6}{c}{C-stat\,/\,d.o.f. = 2485\,/\,2409 (MEG) and 2656\,/\,2609 (HEG)} \\
\hline
\end{tabular}
\end{minipage}
\tablefoot{
For the absorption components (Comps. A to D), the turbulent velocity $\sigma_v$ is ${42\pm10}$~\kms. For the photoionised emission components, $\sigma_v$ is ${35 \pm 20}$~\kms (Comp. E) and ${590 \pm 180}$~\kms (Comp. F). In Comp C, 72\% of the column density is present in the slowest sub-component, and 14\% in each of the other two sub-components.
}
\end{table*}

%
\begin{figure}[!tbp]
\vspace{-0.35cm}
\resizebox{\hsize}{!}{\hspace{-0.3cm}\includegraphics[angle=0]{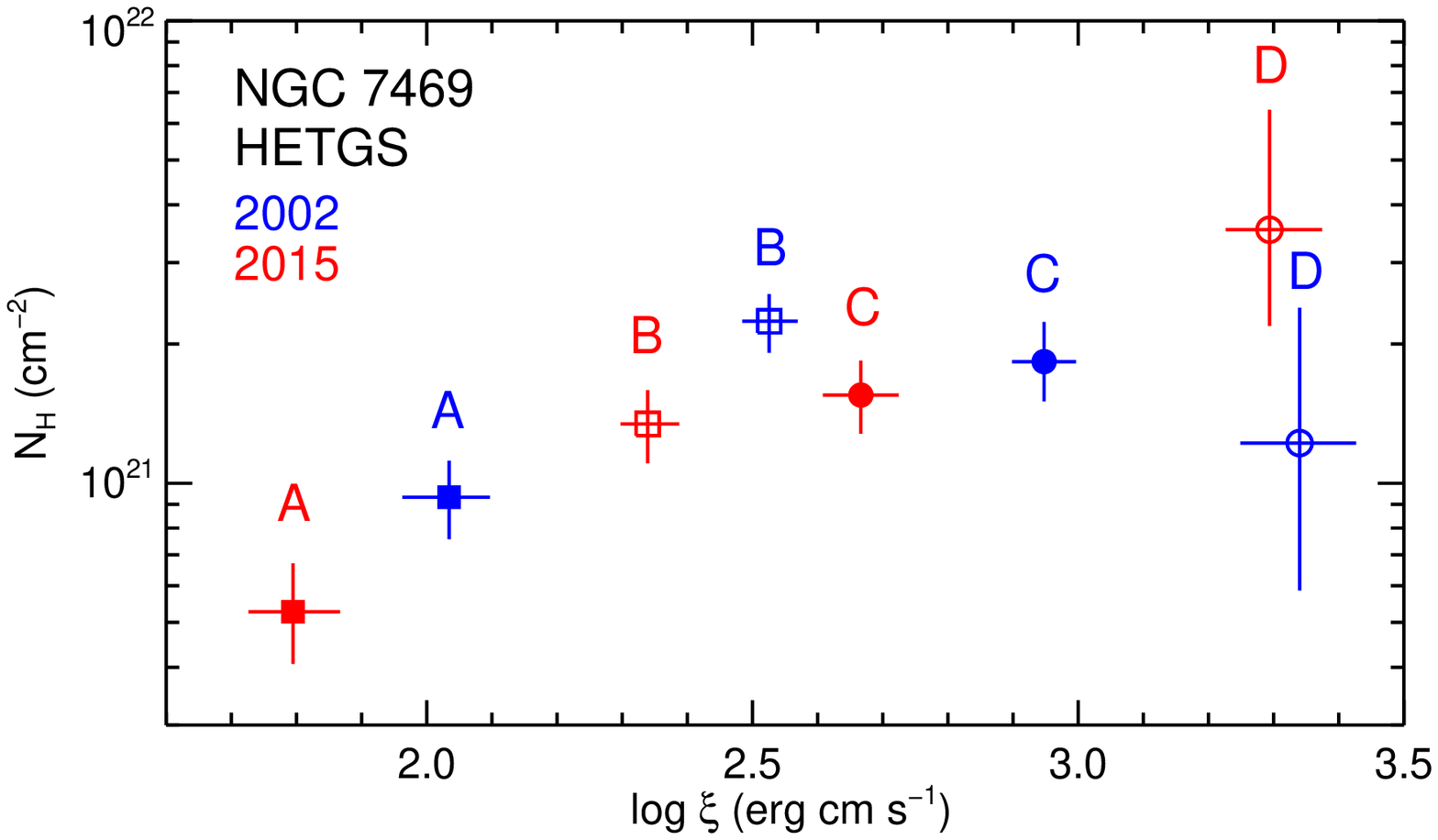}}\vspace{-0.4cm}
\resizebox{\hsize}{!}{\hspace{-0.3cm}\includegraphics[angle=0]{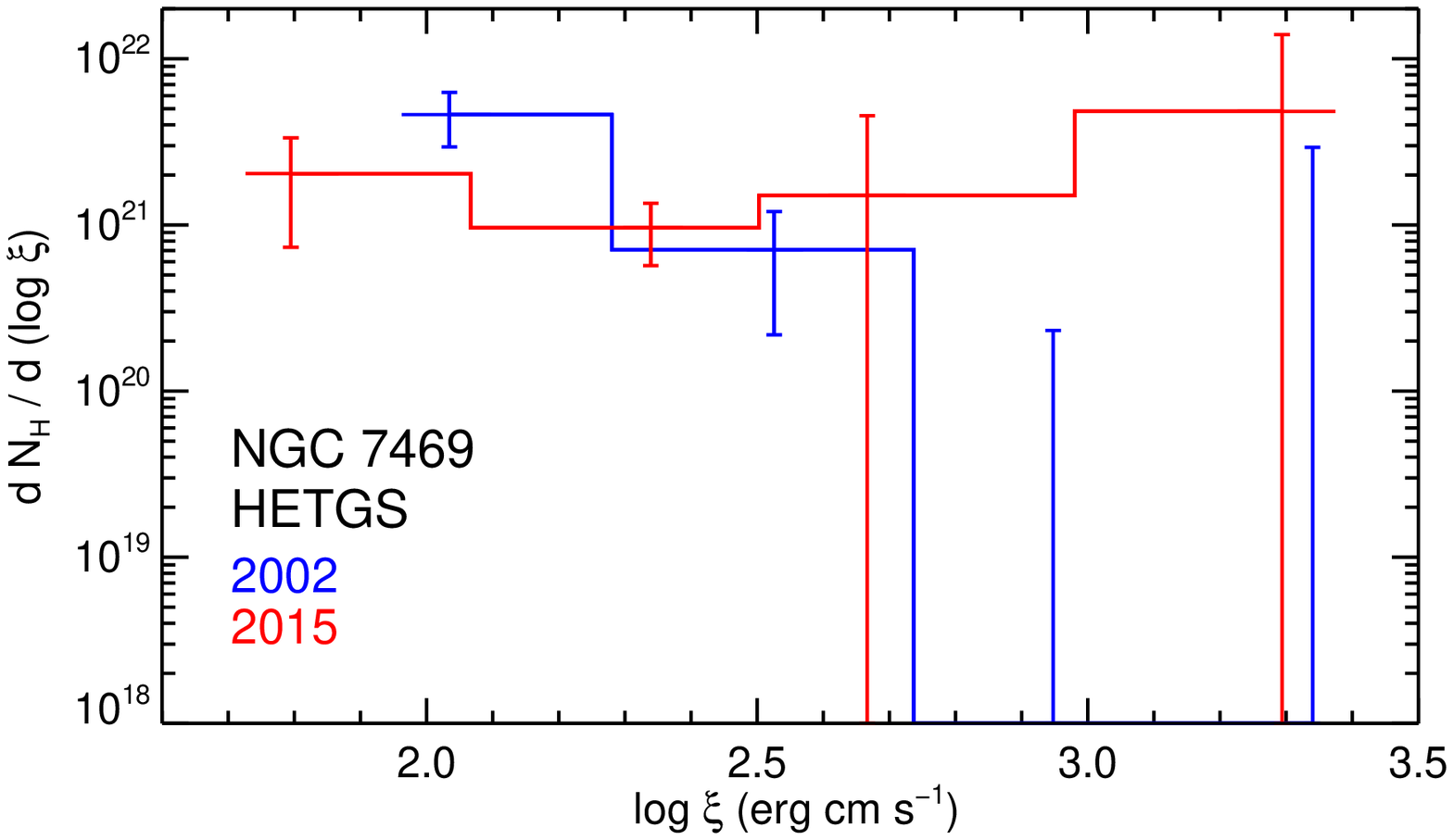}}
\caption{{\it Top panel:} Column density \NH versus the ionisation parameter $\xi$ of the AGN wind absorption components (Comps. A to D) in \ngc, derived for the 2002 (shown in blue) and 2015 (shown in red) \chandra HETGS observations. {\it Bottom panel:} The absorption measure distribution (AMD) corresponding to the components.}
\label{comps_fig}
\end{figure}

The column density \NH, ionisation parameter $\xi$, and outflow velocity $v_{\rm out}$ of each \pion component are fitted. The covering fraction of all our absorption components in our modelling is fixed to unity. The velocity profile of \ion{Ne}{x} Ly$\alpha$ absorption line shows that it consists of three velocity components (Fig. \ref{NeX_fig}), imprinting line absorption at about $-510$, $-1080$, and $-1850$ \kms. About 85\% of the \ion{Ne}{x} column density is present at the lowest velocity, and about 5\% at the highest. These velocities are consistent with those seen for the Ly$\alpha$ lines of the H-like \ion{Mg}{xii} and \ion{Si}{xiv} ions in the HETGS spectrum, and also the velocity components of the \ion{O}{viii} Ly$\alpha$ line detected in the stacked RGS spectrum \citep{Beh17}. The velocities of the X-ray lines are also consistent with the discrete velocities of narrow absorption troughs seen in the high-resolution UV spectra from STIS \citep{Scot05} and COS \citep{Arav17}. Comparable wind velocities in UV and X-rays have also been seen in other AGN (e.g. \object{NGC 3783}, \citealt{Scot14}). This indicates that a single kinematic component of the AGN wind may be associated to multiple ionisation sub-components. One explanation for this is that the kinematic component is a bulk of clumpy gas, which consists of zones with different ionisation states arising from inhomogeneities in density, as well as stratification. Such X-ray and UV absorbing clumps in typical warm-absorber winds (like in NGC 7469) can be explained by photoionised gas being ablated from the torus (see e.g. \citealt{Bals93,Krol95,Krol01}). These clumps may not necessarily be in pressure equilibrium, which in that case are unstable and eventually evaporate. However, they can be replaced by new clumps as they ablate from the torus. So torus wind models can produce density variations with comparable velocities, which would give a range in the observed ionisation parameter of the wind.

In our modelling of the HETGS spectra, we allow Comp. C (which produces most of the aforementioned H-like ions) to have three velocity components. The other components are found to be consistent with each having one outflow velocity. In our modelling we linked the turbulent velocity $\sigma_{v}$ of all the \pion absorption components, in order to fit only one $\sigma_{v}$. This approximation for $\sigma_v$ of the components is sufficient for fitting the detected absorption lines by HETGS. The derived $\sigma_v$ from our global modelling is about 40--50 \kms, which is consistent with the measured $\sigma_v$ of individual UV absorption troughs, ranging between 35 and 60 \kms \citep{Beh17}.

The model transmission spectrum of the four \pion absorption components (Comps. A to D) are displayed in Fig. \ref{trans_fig}. Apart from the absorption lines, the HETGS spectrum of \ngc shows the presence of narrow emission lines, which we fit using the \pion photoionisation model. The \pion model in emission has an additional free parameter which is the emission covering factor ${\Omega\,/\,4\pi}$. In the HETGS spectrum, the Ly$\alpha$ lines of \ion{O}{viii}, \ion{Ne}{x}, and \ion{Mg}{xii} indicate emission at zero velocity (see Fig. \ref{NeX_fig}). Moreover, the \ion{Ne}{ix} and \ion{O}{vii} triplets are also seen through the detection of mostly their forbidden lines. To fit these emission lines we require two \pion components (Comps. E and F). The lower-ionisation component (Comp. E) produces the \ion{Ne}{ix} and \ion{O}{vii} triplets, while the higher-ionisation component (Comp. F) produces the Ly$\alpha$ emission lines of \ion{O}{viii}, \ion{Ne}{x}, and \ion{Mg}{xii}. Our best-fit \pion model to the stacked HETGS spectrum is shown in Fig. \ref{meg_fig}. The best-fit model parameters of the absorption components (Comps. A to D) and emission components (Comps. E and F) are given in Table \ref{pion_table}. For the emission components, the Emission Measure (EM), defined as ${\int n_{\rm e}\, n_{\rm H}\, {\rm d}V}$, is calculated from the fitted parameters of the \pion model (Table \ref{pion_table}). 

We finally applied the stacked \pion model to fit the individual 2002 and 2015 HETGS spectra to look for long-term changes in the photoionised plasma. We freed \NH and $\xi$ of the \pion components, as well as the parameters of the broadband continuum, to fit the 2002 and 2015 data. We find changes in the parameters of the absorption components between the two epochs, while the emission components remain unchanged within errors. The derived \NH and $\xi$ of the absorption \pion components for the 2002 and 2015 observations are shown in Fig. \ref{comps_fig} (top panel). The corresponding absorption measure distribution (AMD), defined as ${{\rm d}\,\NH / {\rm d}\,(\log \xi)}$, is shown in the bottom panel of Fig. \ref{comps_fig}. We have used quadratic Lagrangian interpolation to compute this derivative for the \pion components.

\subsection{Ionisation and thermal state of the photoionised wind}

\begin{figure}[!tbp]
\centering
\vspace{-0.3cm}
\resizebox{\hsize}{!}{\includegraphics[angle=0]{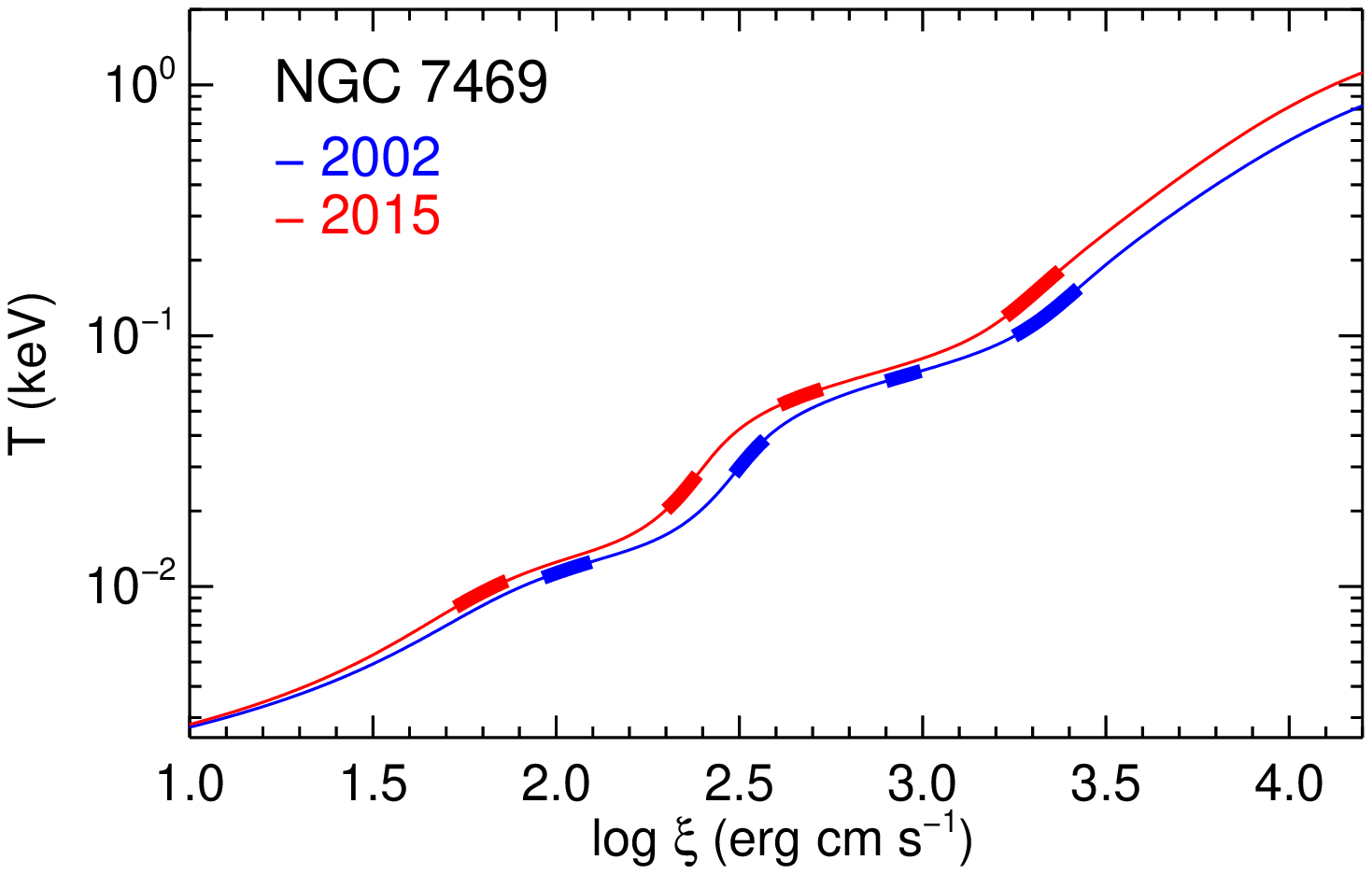}\hspace{0.5cm}}\vspace{-0.3cm}
\resizebox{\hsize}{!}{\includegraphics[angle=0]{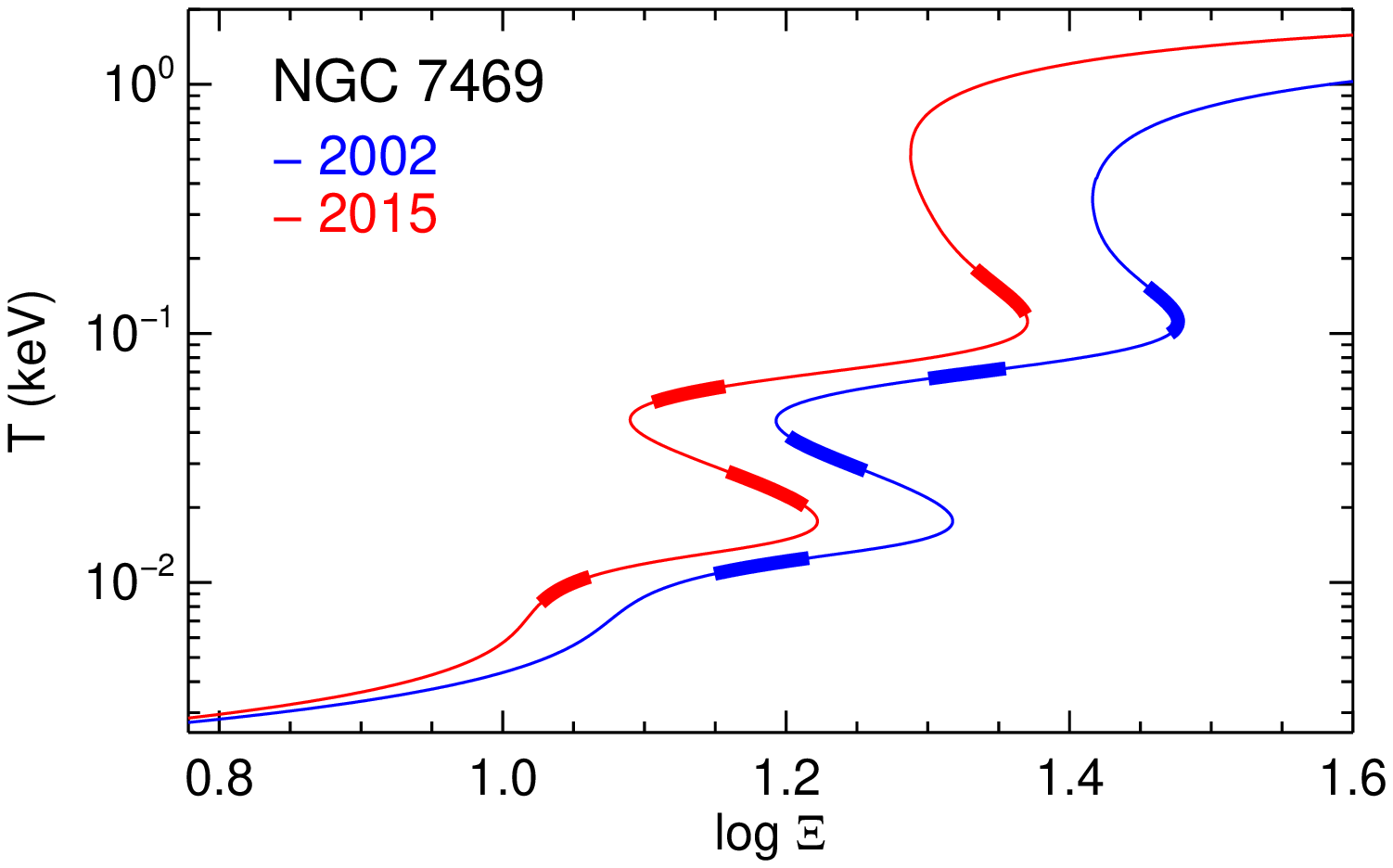}\hspace{0.5cm}}\vspace{-0.2cm}
\caption{Thermal stability curves of photoionised gas in \ngc, showing the gas electron temperature $T$ plotted as a function of the ionisation parameters $\xi$ ({\it top panel}), and the pressure-form of the ionisation parameter $\Xi$ ({\it bottom panel}). The curves are calculated for the 2002 and 2015 SEDs (Fig. \ref{sed_fig}, {\it bottom panel}). The thick stripes on each curve indicate the position of the derived wind absorption components (Comps. A to D) from the best-fit model to the 2002 and 2015 HETGS spectra. For identification, Comp. A has the lowest $T$, and Comp. D the highest.}
\label{stability_fig}
\end{figure}

From our photoionisation and spectral modelling of the AGN wind in \ngc (Sects. \ref{photoionisation_sect} and \ref{spectroscopy_sect}), we find that there are changes in $\xi$ and \NH of the wind absorption components between the 2002 and 2015 observations (Fig. \ref{comps_fig}, top panel). To investigate the origin of these changes, we analyse the ionisation and thermal state of the components, and examine their derived heating and cooling rates. The \pion photoionisation model is used for all our calculations here. The SED determines the ionisation/thermal balance and stability of a plasma in photoionisation equilibrium, which can be thermally unstable in certain regions of the ionisation parameter space. This can be investigated by means of producing thermal stability curves (also called S-curves), which is a plot of the electron temperature $T$ as a function of the pressure form of the ionisation parameter, $\Xi$ \citep{Kro81}. The dimensionless ionisation parameter $\Xi$, is defined as $\Xi  \equiv {F}/{{n_{\rm{H}}\, c\, kT}}$, where $F$ is the flux of the ionising source between 1--1000 Ryd, $k$ is the Boltzmann constant, $T$ is the electron temperature, and $n_{\rm{H}}$ is the hydrogen density. Figure \ref{stability_fig} (top panel) demonstrates the different impact of the 2002 and 2015 SEDs (Fig. \ref{sed_fig}, bottom panel) on the photoionisation equilibrium, and hence the derived temperature $T$ of the plasma. The associated thermal stability curves from the \pion calculations for the 2002 and 2015 SEDs are shown in Fig. \ref{stability_fig} (bottom panel). The position of the wind absorption components derived in Sect. \ref{spectroscopy_sect} are indicated on each curve. 

\begin{figure}[!tbp]
\centering
\vspace{-0.3cm}
\resizebox{\hsize}{!}{\includegraphics[angle=0]{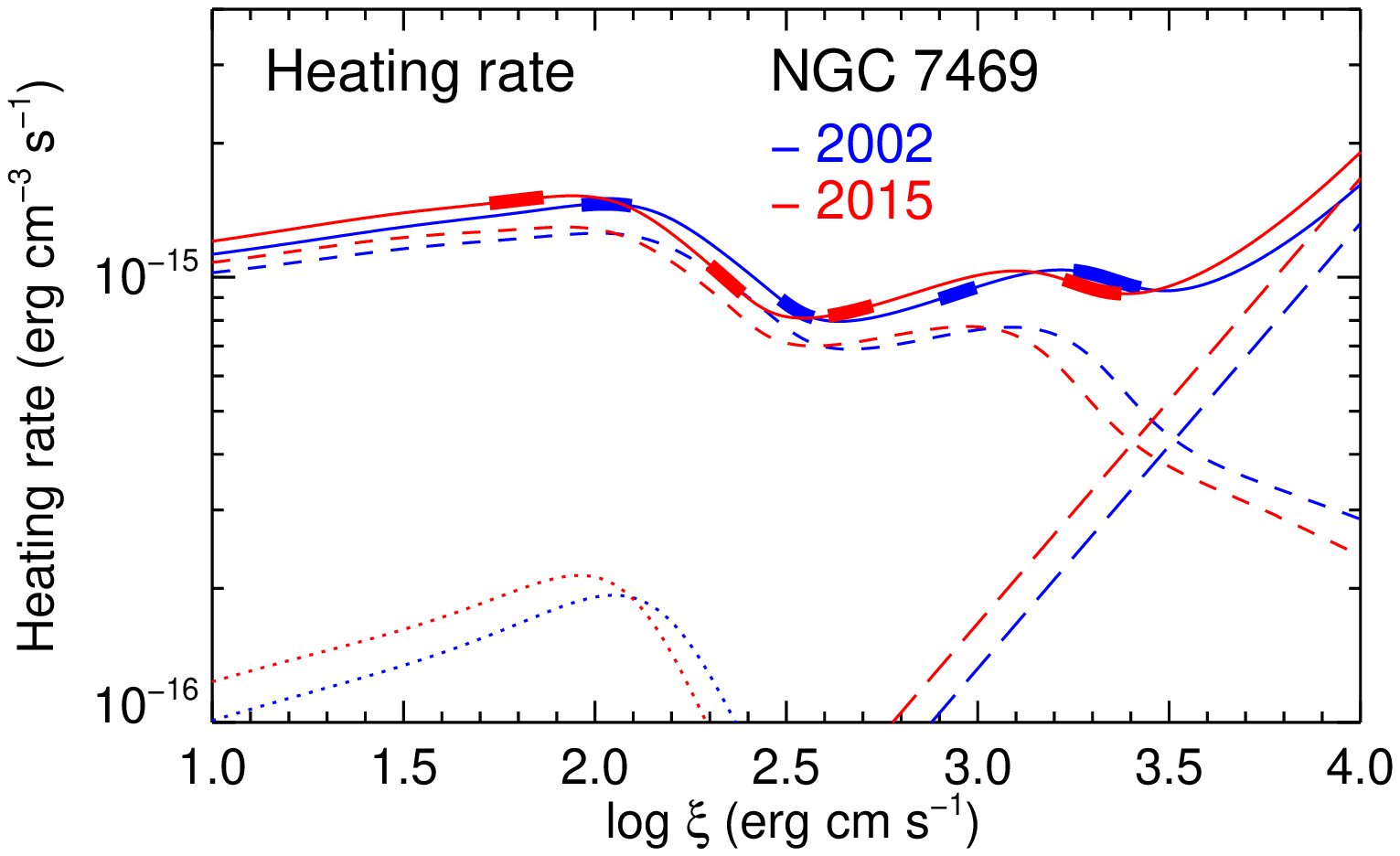}\hspace{0.5cm}}\vspace{-0.3cm}
\resizebox{\hsize}{!}{\includegraphics[angle=0]{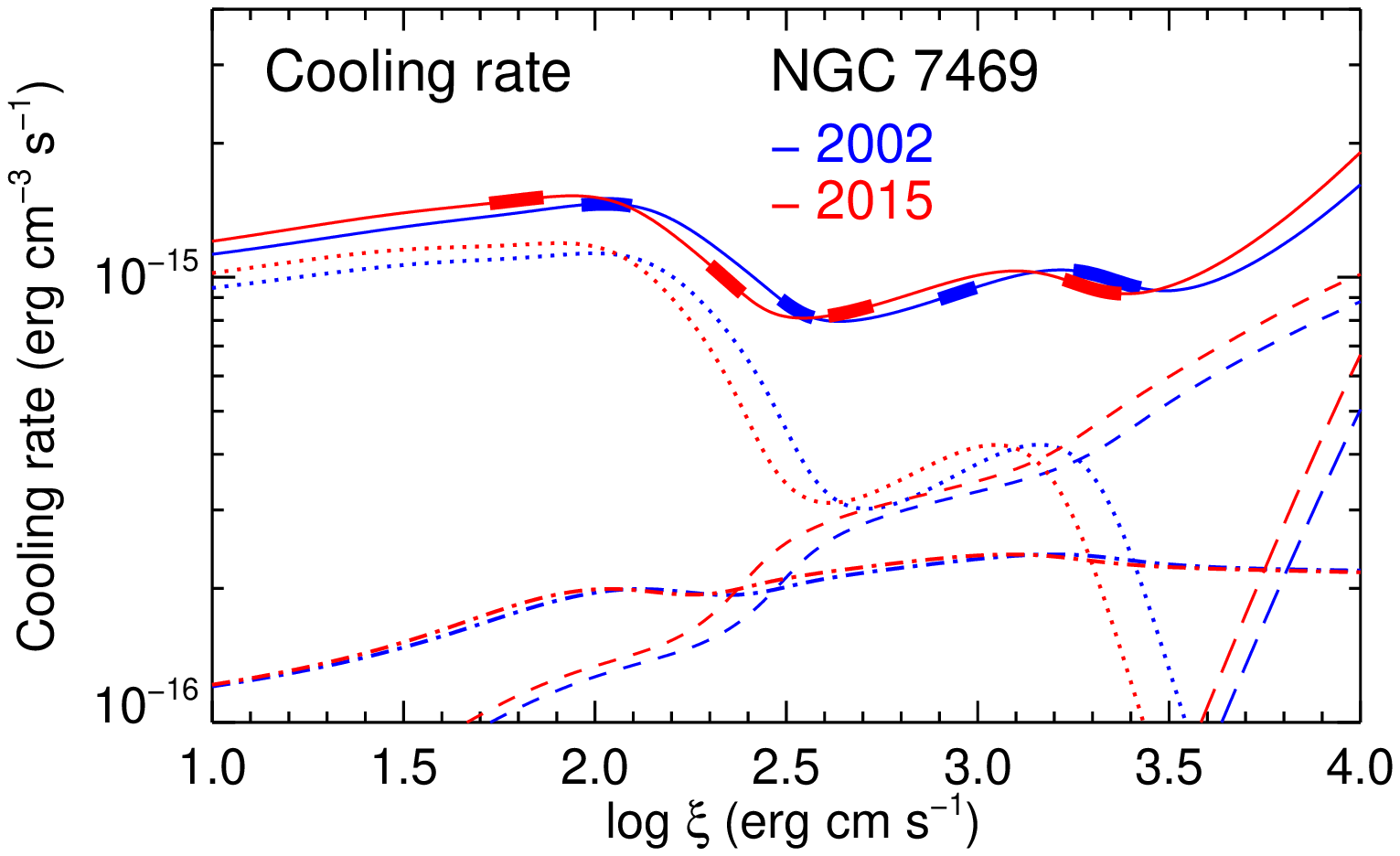}\hspace{0.5cm}}\vspace{-0.2cm}
\caption{Heating rate ({\it top panel}) and cooling rate ({\it bottom panel}) of the photoionised gas in \ngc as a function of the ionisation parameter $\xi$. The contributions by the main processes to the total rates (shown in solid lines) are individually shown. Heating by Compton scattering (long-dashed lines), photoelectrons (dashed lines), and Auger electrons (dotted lines) are displayed in the {\it top panel}. Cooling by inverse Compton scattering (long-dashed lines), Bremsstrahlung (dashed lines), collisional excitation (dotted lines), and recombination (dashed-dotted lines) are displayed in the {\it bottom panel}. The curves are calculated for the 2002 and 2015 SEDs (Fig. \ref{sed_fig}, {\it bottom panel}). The thick stripes on each curve indicate the position of the derived wind absorption components (Comps. A to D) from the best-fit model to the 2002 and 2015 HETGS spectra. For identification, Comp. A has the lowest $\xi$ and Comp. D the highest.}
\label{cooling_fig}
\end{figure}

At photoionisation equilibrium there is a balance between the total heating and cooling in the plasma. The primary heating processes are heating by photoelectrons, Auger electrons, and Compton scattering. The primary cooling processes are collisional excitation, recombination, Bremsstrahlung, and inverse Compton scattering. In Fig. \ref{cooling_fig}, we show the heating and cooling rates of the photoionised plasma in \ngc as a function of $\xi$ for the 2002 and 2015 SEDs, obtained from the \pion calculations. The heating and cooling rates of the derived wind components for the two epochs are also indicated on the curves. The contributions by individual heating and cooling processes are displayed, which enable us to understand how each process acts under the different SEDs, and hence see its impact on the wind components. We discuss these results in Sect. \ref{wind_sect}.

%

%

\section{Diffuse soft X-ray emission in \ngc}
\label{extended_sect}
%

%
\begin{figure}[!tbp]
\centering
\resizebox{\hsize}{!}{\includegraphics[angle=0]{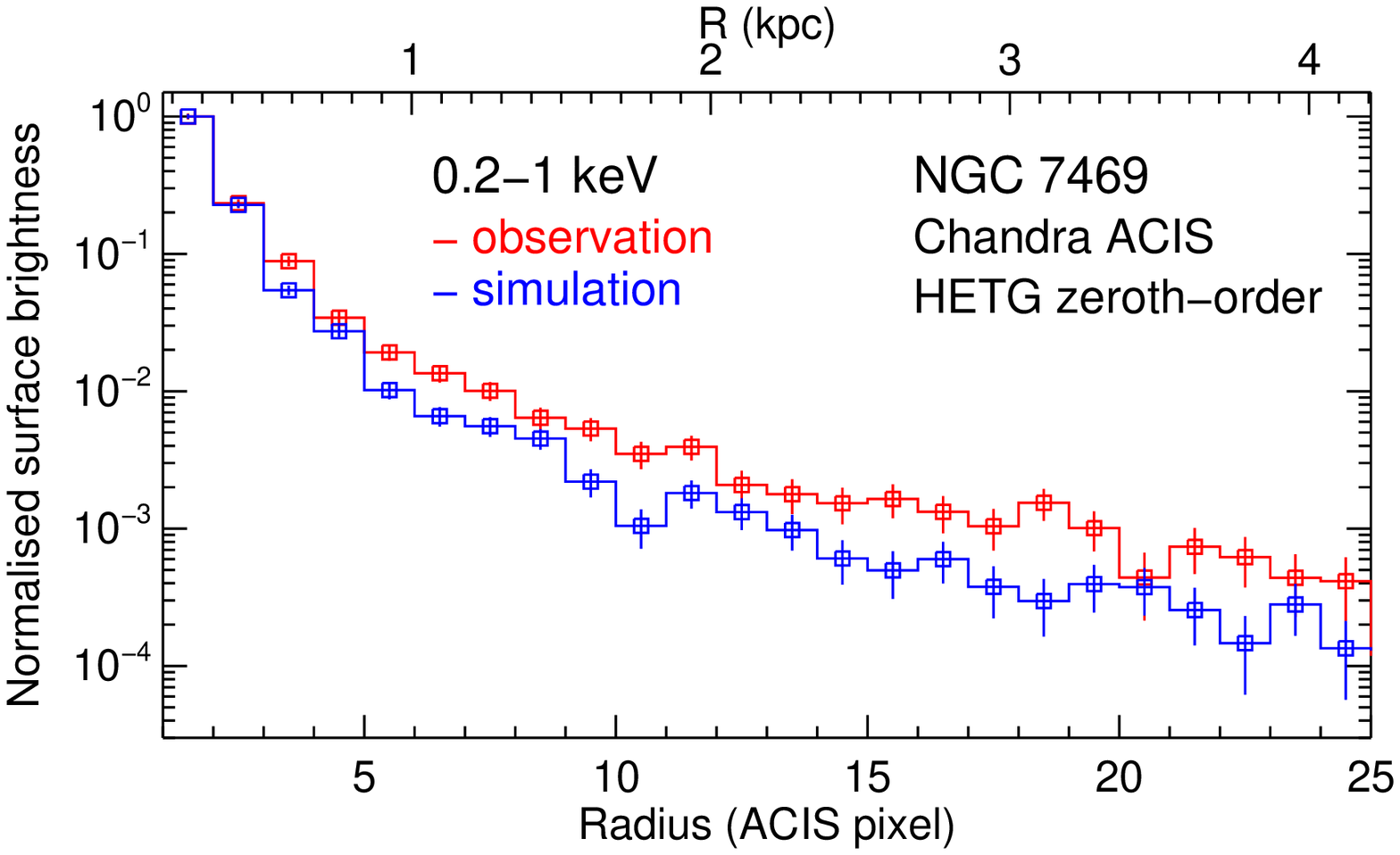}}
\resizebox{\hsize}{!}{\includegraphics[angle=0]{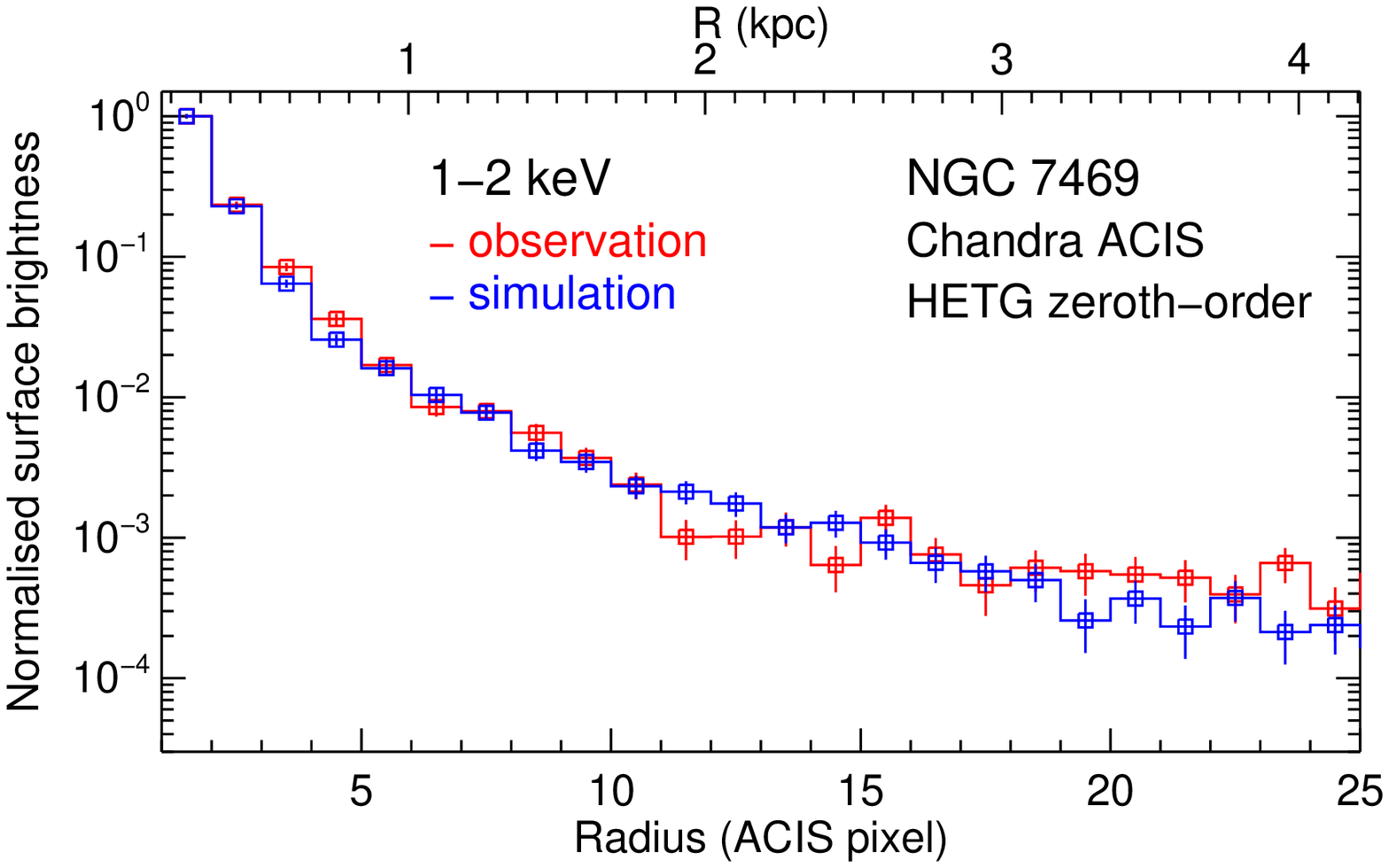}}
\resizebox{\hsize}{!}{\includegraphics[angle=0]{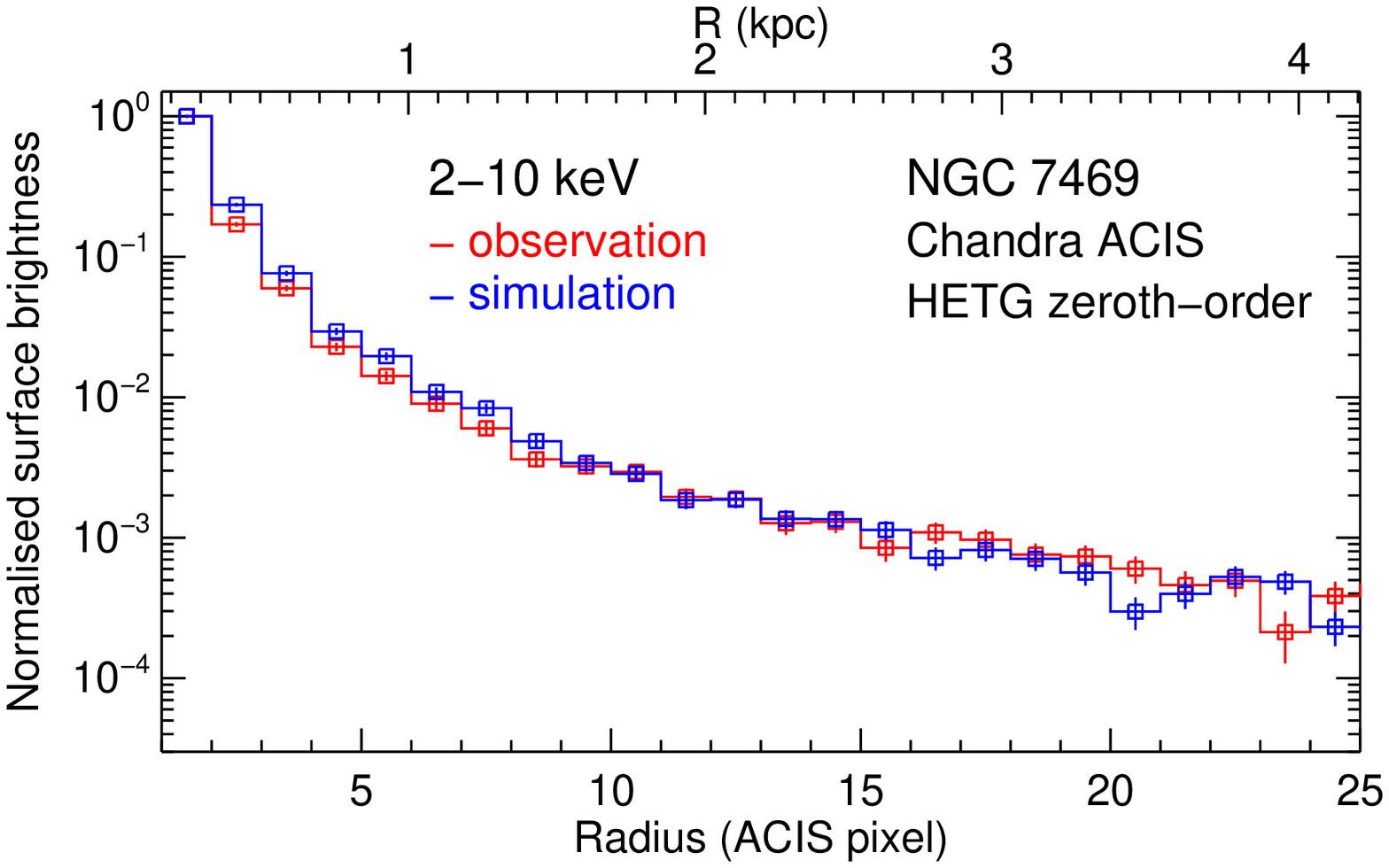}}
\caption{X-ray surface brightness profile of \ngc from a \chandra ACIS HETG zeroth-order observation ({\it shown in red}) and the corresponding PSF simulation ({\it shown in blue}). The results are for \chandra Obs. ID 2956. The profiles are shown for the 0.2--1 keV ({\it top panel}), 1--2 keV ({\it middle panel}), and 2--10 keV ({\it bottom panel}) energy bands. An excess soft X-ray emission (0.2--1 keV) is evident above the PSF in the {\it top panel} from 3 to 25 ACIS pixels (1.5--12$\arcsec$).}
\label{radial_fig}
\end{figure}

As described in Sect. \ref{bulge_sect}, the host galaxy of \ngc exhibits strong starburst activity, which is reported by extensive previous studies from radio to optical/UV energies. Interestingly, there is evidence of X-ray line emission from a CIE plasma in the stacked 640~ks RGS spectrum of \ngc \citep{Beh17}. This \ion{Fe}{xvii} line emission is also marginally detected at $17.38~\AA$ in the first-order HETGS spectrum (Fig. \ref{meg_fig}). As currently the \chandra observatory has the best X-ray spatial resolution (FWHM~${\sim 0.4\arcsec}$), we investigated whether the spatial extent of the X-ray emission from the starburst activity can be detected by \chandra. Since all the \chandra observations of \ngc are taken with the ACIS/HETG gratings, we use the zeroth-order data for this investigation. Here, we obtain the zeroth-order surface brightness radial profile and spectrum of \ngc, and compare with the corresponding \chandra simulations for a point source detection.

For our analysis we used the longest \chandra observation (79~ks), which was taken in 2002 (Obs. ID: 2956), and does not suffer from the low-energy ACIS QE contamination. Using the \chandra~{\tt CIAO} tools we extracted the surface brightness profile from three energy bands: 0.2--1 keV, 1--2 keV, and 2--10 keV. We calculated these using a stack of concentric annuli around the central source, from 1 to 40 pixel radii with increments of 1 pixel. The background was extracted from a source-free region on the same chip (ACIS-S3) using a circle of 40 pixel radius. We then simulated the point spread function (PSF) associated to our particular observation using the Chandra Ray Tracer tool (\chart v2, \citealt{Cart03}). As the \chandra PSF depends on energy, we used the \ngc spectrum at the entrance aperture of HRMA for the \chart simulations. All the simulation parameters are matched to the settings of our observation. The \chart rays were then projected onto the ACIS detector plane to create an event file using the \marx v5.3.2 tool \citep{Davi12}. The image and zeroth-order spectra were then extracted from this simulated event file. As our objective is to simulate a particular observation, we include the effects of pileup in our \marx simulations. Figure \ref{radial_fig} shows the comparison of the X-ray surface brightness profile of \ngc from the \chandra observation and simulation. Interestingly, there is evidence of extended emission in the 0.2--1 keV band at 3--25 ACIS pixel radii (1.5--12$\arcsec$) from the central AGN source. The location of the diffuse soft X-ray emission is consistent with the location of the inner and outer starburst rings seen in IR/optical/UV observations (Sect. \ref{bulge_sect}). Also, the diffuse soft X-ray emission appears to extend beyond the starburst rings, which may be attributed to a starburst wind.

\begin{figure}[!tbp]
\centering
\resizebox{\hsize}{!}{\includegraphics[angle=0]{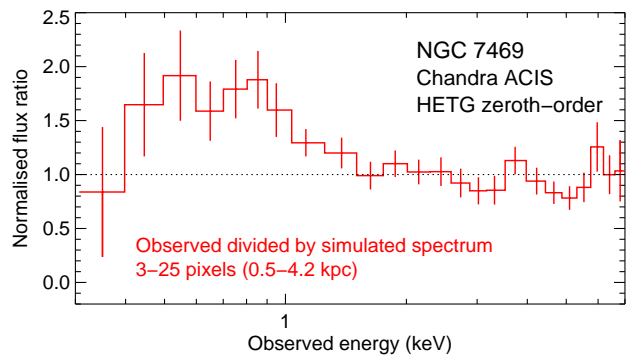}}
\caption{Ratio of the observed over simulated \chandra ACIS HETG zeroth-order spectrum of \ngc in the 3--25 ACIS pixels (1.5--12$\arcsec$). The results are for \chandra Obs. ID 2956.}
\label{zeroth_spec_fig}
\end{figure}

We then extracted the ACIS/HETG zeroth-order spectrum from our observation and simulation using the {\tt specextract} script of {\tt CIAO}. In Fig. \ref{zeroth_spec_fig} we show the ratio of the observed over the simulated spectrum for the 1.5--12$\arcsec$ region. A diffuse excess emission is indeed evident in soft X-rays. We fitted this excess emission with the \cie component reported in Sect. \ref{bulge_sect}, derived from the high-resolution spectroscopy of \ion{Fe}{xvii} coronal lines. The EM of this component was fitted, while its temperature and velocity were kept fixed to the parameter values described earlier in Sect. \ref{bulge_sect} (${T=0.35}$~keV and ${v_{\rm out} = -250}$~\kms). We obtain a good fit with EM of ${6 \pm 3 \times 10^{63}}$~cm$^{-3}$, which is consistent with the original EM found from RGS (${4.0 \pm 1.5 \times 10^{63}}$~cm$^{-3}$, \citealt{Beh17}). The flux of the soft X-ray extended emission over 1.5--12$\arcsec$ is about ${5.4 \times 10^{-13}}$~\ergflux, which matches the flux of the \cie component derived from the first-order spectra. Since the diffuse soft X-ray emission is co-spatial with the nuclear starburst seen at IR/optical/UV energies (e.g. with HST), and both the zeroth-order and first-order spectroscopic modelling results are consistent with the same CIE emission model, we deduce that the diffuse soft X-ray emission originates from the nuclear starburst in \ngc. We further discuss these findings in Sect. \ref{extended_origin_sect}.

%

\section{Discussion}
\label{discussion}

\subsection{Physical structure of the wind in \ngc}
\label{wind_sect}

From our HETGS modelling in Sect. \ref{model_sect}, we determined that the wind in \ngc consists of four photoionisation components (Comps. A to D) in both the 2002 and 2015 epochs. The ionisation parameter $\xi$ of these components appears to change between the 2002 and 2015 epochs (Fig. \ref{comps_fig}, top panel). The $\xi$ is reduced from 2002 to 2015 by a factor of 0.58 in Comp. A, 0.65 in Comp. B, 0.52 in Comp. C, and 0.90 in Comp. D. These changes in $\xi$ are likely related to the change in the SED shape between the two epochs (Fig. \ref{sed_fig}, bottom panel), where the X-ray continuum is higher and the UV continuum is lower in 2015 than in 2002. The consequence of the SED change on the individual heating and cooling processes in the photoionised plasma can be seen in Fig. \ref{cooling_fig}, which causes the changes in $\xi$. 

The absorption measure distribution (AMD), defined as ${{\rm d}\,\NH / {\rm d}\,(\log \xi)}$, is a powerful tool to probe the uncertain physical properties of the AGN wind (see e.g. \citealt{Holc07,Beha09,Adhi15}). It is shown in \citet{Beha09} that the slope of the AMD constrains the density radial profile of the wind. This information derived from X-ray and UV spectra is valuable for matching the observations with the theoretical outflow mechanisms. The AMD shape can be explained by absorbing clouds in photoionisation equilibrium \citep{Roza06,Adhi15,Goos16}. However, a gap is often observed in the middle of the AMD, producing a bi-modal distribution (e.g. \citealt{Beha09}). This gap has been attributed to thermally unstable gas \citep{Adhi15}. In \citet{Pere18} we determined the AMD of \ngc from an ion-by-ion fitting approach using the stacked RGS spectrum of our \xmm campaign. The resulting AMD indicates a dip between $\log \xi$ of about 1 and 2, which is consistent with that previously found from archival data by \citet{Blus07} and \citet{Beha09}. In the present study with HETGS, the AMDs derived for the 2002 and 2015 epochs are shown in Fig. \ref{comps_fig} (bottom panel). As with HETGS we are detecting gas with higher ionisations (${1.8 \lesssim \log \xi \lesssim 3.4}$), the gap in the AMD at lower ionisations is not apparent. By comparing the AMDs in Fig. \ref{comps_fig} (bottom panel), it appears that the 2002 AMD shape drops towards higher ionisations relative to the 2015 AMD. But the associated uncertainties on these AMD points at high ionisations are too large. Therefore, the slopes of the two AMDs are consistent with each other, and with those of the previous studies.

Since the absorption components in \ngc appear to respond to a change in the SED shape between 2002 and 2015, we can put limits on their distance $r$ from the ionising source, using the recombination timescale ${t_{\rm rec}}$. The recombination timescale $t_{\rm rec}$ of an ion recombining from ionisation state ${i+1}$ to $i$ can be written as
\begin{equation}
{t_{\rm rec} =  n_{i}\, /\, (n_{i+1}\, \alpha_{i+1}\, n_{\rm e} - n_{i}\, \alpha_{i}\, n_{\rm e})}, 
\end{equation}
where $n_{\rm e}$ is the electron density, ${n_{i+1}}$ and ${n_{i}}$ are the densities of the ions in state ${i+1}$ and $i$, respectively, and ${\alpha_{i+1}}$ and ${\alpha_{i}}$ are the recombination rate coefficients (${{\rm cm}^{3}~{\rm s}^{-1}}$) from state ${i + 1}$ to $i$, and ${i}$ to ${i-1}$, respectively. We computed ${t_{\rm rec}}$ for the most relevant and dominant Ne ion in each component, which corresponds to recombination from \ion{Ne}{x} to \ion{Ne}{ix} for Comps. B to D, and recombination from \ion{Ne}{ix} to \ion{Ne}{viii} for Comp. A. Since ${t_{\rm rec}}$, which is dependent on the gas density, needs to be shorter than the spacing between the 2002 and 2015 epochs, this can be used to put constraints on $n_{\rm H}$. Hence, from the definition of $\xi$, the distance ${r = \sqrt{L / \xi\, n_{\rm H}}}$ can be calculated. Using ${t_{\rm rec} < 13}$~years, we derive the location of the absorption components in \ngc to be at ${r < 80}$~pc (Comp. A), ${r < 29}$~pc (Comp. B), and ${r < 31}$~pc (Comps. C and D). 

%
\begin{table}[!tbp]
\begin{minipage}[t]{\hsize}
\setlength{\extrarowheight}{3pt}
\caption{The recombination ${t_{\rm rec}}$, cooling ${t_{\rm cool}}$, dynamical ${t_{\rm dyn}}$, and sound crossing ${t_{\rm s}}$ timescales for the wind components in \ngc. These density-dependent timescales are given in units of $n_{4}^{-1}$ days or years, where $n_{4}$ represents ${n_{\rm H} = 10^{4}}$~cm$^{-3}$.}
\label{timescale_table}
\centering
\small
\renewcommand{\footnoterule}{}
\begin{tabular}{c | c c c c}
\hline \hline
Wind		& ${t_{\rm rec}}$		& ${t_{\rm cool}}$		& ${t_{\rm dyn}}$				& ${t_{\rm s}}$ 	\\
Comp.	& ($n_{4}^{-1}$ days) 	& ($n_{4}^{-1}$ years)	& ($n_{4}^{-1}$ years)			& ($n_{4}^{-1}$ years)		\\
\hline
A			& 14				& 0.1					& 44							& 452				\\
B			& 32				& 0.5					& 124						& 674				\\
C			& 12				& 1.4					& 55							& 290				\\
D			& 3.6	 			& 3.0					& 196						& 361				\\
\hline
\end{tabular}
\end{minipage}
\end{table}

Although there are changes in $\xi$ between the 2002 and 2015 observations, the \NH sum of the four absorption components remains effectively unchanged between the two epochs at about 6 to 7~${\times 10^{21}}$~\cm. So while changes in the SED cause a change in the distribution of \NH over different ionisations, the total \NH is still conserved. This suggests that the same absorbing wind is likely present in our line of sight in both epochs. Also, the wind preserves its geometrical configuration despite changes in the ionising SED. Similarly, we find that for the main ions detected in X-rays, the sum of their ionic column densities over all four components, remains in practice unchanged. For example, from our modelling the total column density of \ion{Ne}{x} is ${1.9 \times 10^{17}}$~\cm in 2002 and ${1.5 \times 10^{17}}$~\cm in 2015. Also, the total column density of \ion{O}{viii} is ${5.0 \times 10^{17}}$~\cm in 2002 and ${4.0 \times 10^{17}}$~\cm in 2015. These are consistent with the RGS results of \cite{Pere18}, in which the measured ionic column densities (resulting from all components) are found not to significantly vary between the individual \xmm observations of our 2015 campaign.

According to our thermal stability analysis of photoionised gas in \ngc (Fig. \ref{stability_fig}), two of the four \pion components are thermally unstable (Comps. B and D). As explained by \citet{Kall10}, the existence of thermal instability depends on the validity of the assumption of thermal and ionisation equilibrium. For thermal and ionisation equilibrium to be true, the cooling timescale $t_{\rm cool}$, and the recombination timescale $t_{\rm rec}$, need to be shorter than the dynamical timescale $t_{\rm dyn}$ and the sound-crossing timescale $t_{\rm s}$. The cooling timescale ${t_{\rm cool} = P_{\rm gas} / n_{\rm tot}^2\, \Lambda}$, where gas pressure ${P_{\rm gas} = 3\, n_{\rm tot}\, k\, T / 2}$, $n_{\rm tot}$ is total particle number density of the gas, and $\Lambda$ is the cooling rate coefficient in erg~cm$^3$~s$^{-1}$. Hence, $n_{\rm tot}^2\, \Lambda$ is the total cooling rate of the gas in erg~cm$^{-3}$~s$^{-1}$ (Fig. \ref{cooling_fig}). From photoionisation modelling, $P_{\rm gas}$ and $\Lambda$ are obtained, so $t_{\rm cool}$ can be calculated for a given density. The dynamical timescale ${t_{\rm dyn} = \Delta r / v_{\rm out}}$, where $\Delta r$ is the thickness of the flowing shell ($\sim \NH / n_{\rm H}$) and $v_{\rm out}$ is the outflow velocity of the gas. The sound-crossing timescale ${t_{\rm s} = \Delta r / c_{\rm s}}$, where the speed of sound ${c_{\rm s} = \sqrt{P_{\rm gas} / \rho}}$, and $\rho$ is the gas mass density. These timescales, derived for the wind components in \ngc, are given in Table \ref{timescale_table}.

We find that ${t_{\rm rec} < t_{\rm cool} < t_{\rm dyn} < t_{\rm s}}$ for each absorption component in \ngc. The $t_{\rm rec}$ of our components ranges from 4 to 32~$n_{4}^{-1}$ days, while $t_{\rm cool}$ ranges from 0.1 to 3~$n_{4}^{-1}$ year, where $n_{4}$ represents ${n_{\rm H} = 10^{4}}$~cm$^{-3}$. On the other hand, $t_{\rm dyn}$ ranges from 44 to 196 $n_{4}^{-1}$ years, while $t_{\rm s}$ ranges from 290 to 674 $n_{4}^{-1}$ years. Thus, the criteria for thermal and ionisation equilibrium are satisfied for all the components, including the thermally unstable Comps. B and D. Thermally unstable gas would heat or cool to reach the stable regions of the S-curves (Fig. \ref{stability_fig}, bottom panel). The timescale for this departure depends on $t_{\rm cool}$, which determines how quickly gas can respond to changes in the environment. Since in both the 2002 and 2015 epochs, Comps. B and D reside on the unstable branches of the S-curves, $t_{\rm cool}$ for these components should be more than 13 years, otherwise the gas would have departed from its unstable temperature by the time of the 2015 observation. This information can be used to put limits on $n_{H}$ and $r$. We find $t_{\rm cool} > 13$~years corresponds to ${n_{\rm H} < 405}$~cm$^{-3}$ and ${r > 12}$~pc for Comp. B, and ${n_{\rm H} < 2295}$~cm$^{-3}$ and ${r > 2}$~pc for Comp. D. By combining these $r$ limits from $t_{\rm cool}$, with the ones calculated earlier from $t_{\rm rec}$, we obtain ${12 < r < 29}$~pc (Comp. B) and ${2 < r < 31}$~pc (Comp. D). This suggests that the location of Comp. D, which is the most ionised component of the AMD (Fig. \ref{comps_fig}), is likely the closest to the central source.

From the IR reverberation study of the AGN torus, \citet{Suga06} finds that the inner radius of the torus in \ngc is at 65--87 light days (0.05--0.07 pc). The results of our photoionisation modelling of the wind in \ngc, and the location of the components, are consistent with a wind originating from the AGN torus as previously suggested by \citet{Scot05} and \citet{Blus07}. Different explanations for the driving mechanism of such a torus wind have been reported in the literature, such as a thermally-driven wind from the torus (e.g. \citealt{Krol01}). Another viable explanation is that IR radiation from the torus can drive away the wind (e.g. \citealt{Doro08a,Doro12,Doro16}). Also, the IR radiation pressure on dust grains can boost such winds from the torus \citep{Doro11}. Furthermore, a combination of radiative and magnetic driving from the torus could be in play (e.g. \citealt{Keat12}). Apart from uncertainties in the theoretical modelling of AGN winds, the key parameters of the wind derived from observations (such as the AMD, density, distance, and turbulence) have also large uncertainties associated to them. This makes strong association to one particular model challenging. The upcoming {\it Athena} X-ray observatory, with its X-IFU microcalorimeter \citep{Barr16}, will provide a breakthrough in constraining the uncertain parameters of the wind. In particular it will enable us to accurately probe the high ionisation regions of the wind, which is currently the most uncertain, and yet potentially important, part of the AGN winds.

\subsection{Diffuse X-ray emission from the starburst of \ngc}
\label{extended_origin_sect}

The well-known nuclear starburst ring in \ngc has been observed over different energies from radio to optical/UV, such as studies by \citet{Diaz07} with HST, \citet{Perr09} with VLA, and \citet{Izum15} with ALMA. In Sect. \ref{extended_sect} we analysed the ACIS/HETG zeroth-order data of \ngc to determine the spatial extent of the X-ray emission in \ngc. We found a diffuse soft X-ray component over 1.5--12$\arcsec$ from the central AGN source, which is co-spatial with the nuclear starburst in \ngc. This diffuse X-ray component is consistent with CIE emission from a plasma with a temperature of 0.2--0.3~keV associated to the star formation activity of \ngc. 

The luminosity of this diffuse CIE emission enables us to estimate the star formation rate (SFR) in \ngc, which can be then compared with SFR estimates from the far infrared (FIR) emission (e.g. \citealt{Kenn98}). An extensive study of a sample of 29 nearby star-forming galaxies by \citet{Mine12b} finds that the apparent luminosity of the diffuse emission in the 0.5--2 keV band ($L_{\rm X}^{\rm diff}$) linearly correlates with the star formation rate (SFR) according to: ${{L_{\rm X}^{\rm diff}} (\ergs) \approx 8.3 \times 10^{38}}$ SFR $(M_{\odot}\, {\rm yr}^{-1})$. In the case of \ngc, we derive ${L_{\rm X}^{\rm diff} = 5 \pm 3 \times 10^{40}}$~\ergs following our modelling in Sect. \ref{extended_sect}. Thus, this $L_{\rm X}^{\rm diff}$ implies that the SFR in \ngc is $60 \pm 35$ $M_{\odot}~{\rm yr}^{-1}$.

The FIR (40--400 $\mu$m) luminosity of \ngc is obtained by \citet{Sand03}: ${L_{\rm FIR} = 2.51 \times 10^{11}}$~$L_{\odot}$, where the solar bolometric luminosity ${L_{\odot} = 3.83 \times 10^{33}}$~\ergs. From the $L_{\rm FIR}$--SFR relation of \citet{Kenn98}, one finds that the SFR in \ngc is about 44~$M_{\odot}~{\rm yr}^{-1}$. Moreover, \citet{Pere11} derive the SFR to be about $48$~$M_{\odot}~{\rm yr}^{-1}$ in \ngc, based on the \spitzer/MIPS 24$\mu$m luminosity and the SFR relation of \citet{Riek09}. Therefore, the SFR derived from our diffuse soft X-ray measurements are consistent with those found from $L_{\rm FIR}$--SFR and $L_{\rm X}^{\rm diff}$--SFR relations.

The diffuse soft X-rays discovered in \ngc are unlikely to be associated to the photoionised X-ray emission in \ngc. In Sect. \ref{spectroscopy_sect} we determined two \pion photoionised components (Comps. E and F) that fit the observed narrow X-ray emission lines in the HETGS first-order spectrum (e.g. the \ion{O}{viii} and \ion{Ne}{x} Ly$\alpha$ lines, and the \ion{O}{vii} and \ion{Ne}{ix} triplets). The 0.5--2 keV luminosities of these components are ${8.9 \times 10^{40}}$~\ergs (Comp. E) and ${1.6 \times 10^{41}}$~\ergs (Comp. F). The total luminosity of these two components is significantly higher than that of the diffuse component (${5 \times 10^{40}}$~\ergs), although the luminosity of Comp. E is comparable to that of the diffuse component. However, photoionised emission in \ngc is most likely localised to the inner regions of the nucleus since the size of the NLR from HST [\ion{O}{iii}] images is found to be $< 180$~pc \citep{Mull11}. This is extremely compact compared to the extent of the diffuse soft X-ray component (0.5--4.2 kpc). The X-ray counterpart of the NLR, likely originating from a region with a comparable size to the NLR or smaller, cannot be properly resolved in X-rays as the 0.4$\arcsec$ FWHM of \chandra equates to about 140~pc for \ngc.

\section{Conclusions}
\label{conclusions}

In this investigation we determined the SED of \ngc at two epochs (2002 and 2015) by taking into account all non-intrinsic optical-UV-X-ray processes in our line of sight; studied the AGN wind and its long-term variability through photoionisation modelling and \chandra HETGS spectroscopy; and analysed the spatial extent of the X-ray emission in \ngc. From the findings of our investigation we conclude the following:

\begin{enumerate}
\item There are long-term changes in the AGN continuum shape of \ngc, with the 2015 observation having lower UV flux and higher X-ray flux than the 2002 observation. We find that the broadband continuum of \ngc consists of an optical/UV thermal component from the disk, which is modified by warm Comptonisation to produce the soft X-ray excess. The primary X-ray power-law has a photon index ${\Gamma \approx 1.91}$ in both epochs, but is brighter in 2015 than in 2002. 
\\
\item The \FeKa line in the HETGS spectrum consists of a broad component (${\sigma_v \approx 2700}$~\kms) and an unresolved component (${\sigma_v < 460}$~\kms). The flux and parameters of the reflection model for the \FeKa line are found to be unchanged in both the 2002 and 2015 epochs.
\\
\item The AGN wind in \ngc consists of four photoionised absorption components, with $\log \xi$ ranging from 1.8 to 3.3. The outflow velocity of these components ranges from $-400$ to $-1800$~\kms. The narrow soft X-ray emission lines are modelled with two other photoionisation components. From the long-term changes in the ionisation state of the wind components, we derive limits on their location. For two of the four components, which are found to be thermally unstable in both epochs, we obtain ${2 < r < 31}$~pc and ${12 < r < 29}$~pc using the cooling and recombination timescales. For the other two thermally stable components, we obtain ${r < 31}$~pc and ${r < 80}$~pc from the recombination timescale. The results of our modelling are consistent with a clumpy, thermally-driven, torus wind in \ngc.
\\
\item The analysis of \chandra ACIS/HETG zeroth-order data reveals that the soft X-ray emission is spatially extended in \ngc at 1.5--12$\arcsec$ radii from the central AGN source. This diffuse soft X-ray emission is co-spatial with the nuclear starburst seen in the infrared/optical/UV observations. Our spectral modelling of this diffuse X-ray emission shows that it is consistent with emission from a CIE plasma, produced by the strong starburst activity in \ngc. 
\end{enumerate}
\begin{acknowledgements}

This research has made use of data obtained from the {\it Chandra} Data Archive, and software provided by the {\it Chandra} X-ray Center (CXC). SRON is supported financially by NWO, the Netherlands Organization for Scientific Research. The research at the Technion is supported by the I-CORE program of the Planning and Budgeting Committee (grant number 1937/12). EB received funding from the European Union's Horizon 2020 research and innovation programme under the Marie Sklodowska-Curie grant agreement no. 655324. BDM acknowledges support by the Polish National Science Center grant Polonez 2016/21/P/ST9/04025. POP acknowledges support from the CNES and French PNHE. We thank M. Bentz for providing us with the host galaxy flux measurement in \ngc, and the anonymous referee for the useful comments.

\end{acknowledgements}


\end{document}